\documentclass[numberedappendix]{emulateapj}
\usepackage{natbib}

\begin{document}

\setlength\tabcolsep{0.2pt}

\title{Evolution of Quiescent and Star-Forming Galaxies Since $z\sim1.5$ as a Function of Their Velocity Dispersions}

\author{Rachel Bezanson\altaffilmark{1}, Pieter van Dokkum\altaffilmark{1}, Marijn Franx\altaffilmark{2}}

\altaffiltext{1}{Department of Astronomy, Yale University, New Haven, CT 06520-8101}
\altaffiltext{2}{Sterrewacht Leiden, Leiden University, NL-2300 RA Leiden, Netherlands}

\shorttitle{Evolution of Quiescent and Star-Forming Galaxies Since $z\sim1.5$}

\newcommand{\unit}[1]{\ensuremath{\, \mathrm{#1}}}

\begin{abstract}
We measure stellar masses and structural parameters for 5,500 quiescent and 20,000 star-forming galaxies at $0.3<z\leq1.5$ in the Newfirm Medium Band Survey COSMOS and UKIDSS UDS fields.  We combine these measurements to infer velocity dispersions and determine how the number density of galaxies at fixed inferred dispersion, or the Velocity Dispersion Function (VDF), evolves with time for each population.  We show that the number of galaxies with high velocity dispersions appears to be surprisingly stable with time, regardless of their star formation history.  Furthermore, the overall VDF for star-forming galaxies is constant with redshift, extending down to the lowest velocity dispersions probed by this study.  The only galaxy population showing strong evolution are quiescent galaxies with low inferred dispersions, whose number density increases by a factor of $\sim4$ since $z=1.5$.  This build-up leads to an evolution in the quiescent fraction of galaxies such that the threshold dispersion above which quiescent galaxies dominate the counts moves to lower velocity dispersion with time.  We show that our results are qualitatively consistent with a simple model in which star-forming galaxies quench and are added to the quiescent population.  In order to compensate for the migration into the quiescent population, the velocity dispersions of star-forming galaxies must increase, with a rate that increases with dispersion.

\end{abstract}


\section{Introduction}

The study of galaxy evolution is often focused on identifying the primary properties which drive growth and eventually determine the properties of any given galaxy.  Initial studies emphasized the importance of galaxy luminosity \citep[e.g.][]{bower:92}.  Many recent studies highlight the importance of stellar mass in determining both the star formation history and the black hole growth \citep[e.g.][]{bundy:06,peng:10}.  However, hints that stellar mass cannot be the sole driver of galaxy evolution remain - for example, at intermediate masses both red and blue galaxies exist \citep[e.g.][]{strateva:01,blanton:03, kauffmann:03}.  Increasing evidence points to stellar velocity dispersion, or some combination of stellar mass and size, as a more fundamental property of a galaxy; it is connected to the galaxy's stellar population, the dark matter halo in which it resides and the super-massive black hole (SMBH) at its center.  Velocity dispersion is most notably included as a key parameter in the Fundamental Plane for elliptical galaxies \citep[e.g.][]{djorgovski:87,bernardifp:03} and in the \citet{magorrian:98} relation between $\sigma_{bulge}$ and $M_{\bullet}$.  Recently, striking correlations between velocity dispersion and other galaxy properties such as color and star formation rates (SFRs) have been examined \citep[e.g.][]{franx:08,bell:12,dokkum:11}.  Furthermore, \citet{wake:12a} demonstrated using clustering analysis of the SDSS that the stellar velocity dispersion of a massive galaxy is more strongly related to the halo properties than even stellar or dynamical mass.

In this context, it is important to create a census of galaxies as a function of velocity dispersion and explore how it has evolved over time, as it is directly linked to the evolution of many facets of galaxy properties.  The number density of galaxies at a given velocity dispersion, or the Velocity Dispersion Function (VDF) has been studied extensively in the SDSS \citep[e.g.][]{sheth:03,mitchell:05,choi:07,shankar:09,bernardi:10} both for elliptical galaxies, for which velocity dispersion measurements are most commonly used, as well as other subgroups and the entire galaxy population.  \citet{bezanson:11} used inferred velocity dispersions to demonstrate that the evolution of the VDF out to $z\sim1.5$ is differential - with a build-up of low dispersion galaxies and little evolution in the number of galaxies with high velocity dispersions in time, which is qualitatively consistent with the expected evolution of the VDF due to galaxy ages \citep{shankar:09} but somewhat different from results based on lensing surveys \citep{chae:10}.

Here we extend this work by examining the number densities of star-forming and quiescent galaxies separately, with two motivations.  First, for the overall population of galaxies, velocity dispersion (or compactness) correlates with quiescence, such that velocity dispersion is anti-correlated with star-formation rates \citep{franx:08,bell:12,wake:12b}.  Therefore we expect the distribution as a function of velocity dispersion to be quite different for each type of galaxy. And second, for individual galaxies, velocity dispersion might be a relatively stable property, as opposed to other properties such as luminosity, mass and even sizes.  However velocity dispersion must be able to grow with time, at least initially.  Based on simple Virial arguments, increasing velocity dispersion requires increasing the mass within a fixed radius and therefore some sort of dissipative processes \citep[e.g.][]{hopkins:09scaling}, which would mainly impact star-forming galaxies.  On the other hand, it has been suggested that velocity dispersions could decrease via a process such as minor merging which could puff up galaxies without adding a significant amount of mass \citep[e.g.][]{bezanson:09,naab:09,oser:11}. Although minor merging could be influential for all types of galaxies, this would be more important in denser regions or galaxies embedded in more massive dark matter halos.  

We present the data used in this analysis in \S2 and investigate the evolution in the number density of galaxies as a function of velocity dispersion and surface density in \S3.  \S4 describes a simple model which reproduces the qualitative evolution of the VDF.  Finally we discuss these results and some of the implications for galaxy formation in the context of dark matter haloes and the growth of SMBHs in \S5.  Throughout this work, we assume $H_0=70\rm{\,km/s\,Mpc^{-1}},\Omega_M=0.3\,\&\,\Omega_{\Lambda}=0.7$.

\section{Data}

\begin{figure*}[!!t]
  \centering
  \epsscale{0.8}
  \plotone{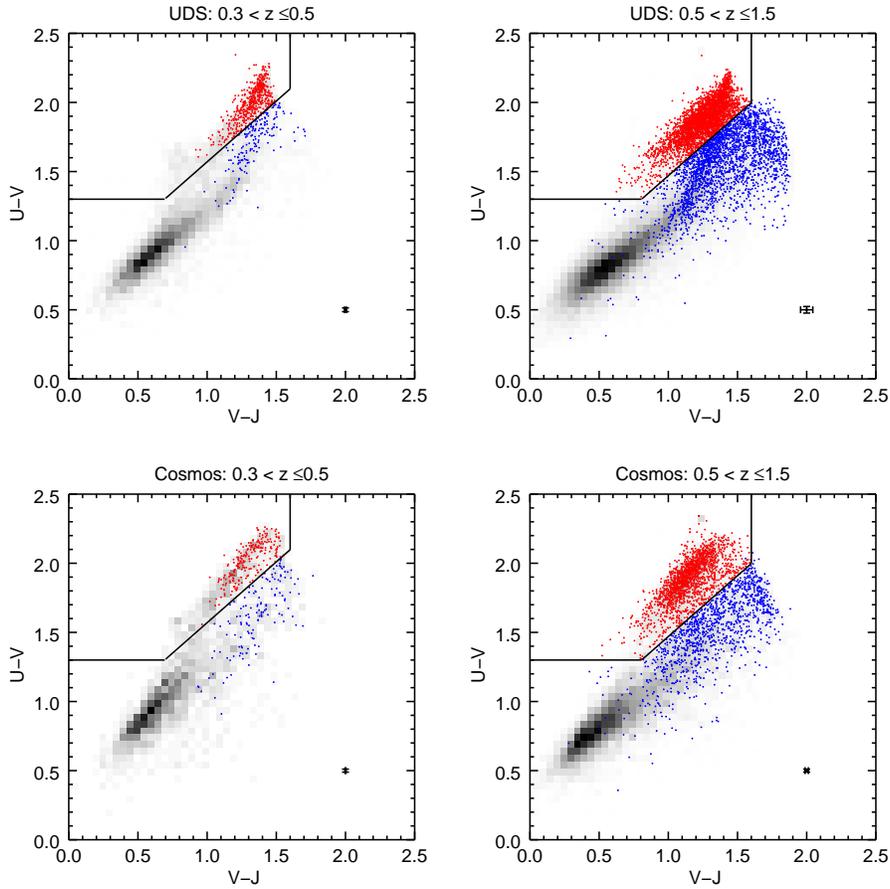}
    \caption{Rest-frame U-V and V-J colors selection diagrams for galaxies in the UDS and NMBS Cosmos fields for two redshift ranges.  Grey scale boxes indicate the overall distribution of galaxies in the redshift range and colored points highlight galaxies above the $\sigma_{inf}\geq\,100\unit{km/s}$ threshold: red and blue points mark quiescent and star-forming galaxies.  Median errors in rest-frame colors for selected galaxies are shown in the lower right corner of each panel.}
  \label{fig:UVJ}
\end{figure*}

\subsection{Redshifts,Stellar Masses, Sizes and Inferred Velocity Dispersions}

For this work we use photometric measurements of galaxy stellar mass, effective radii and S\'{e}rsic index to estimate inferred velocity dispersions for galaxies in the NMBS COSMOS \citep{whitaker:11} and UKIDSS UDS fields \citep{williams:09} and study the redshift evolution of the VDF.   Inferred velocity dispersion has been shown to correlate with measured velocity dispersions in the SDSS, especially when incorporating the galaxy structure in the form of the \citet{sersic} n parameter \citep{taylor:10,bezanson:11}.  Details of the technique, including the calibration of this relation using the SDSS, can be found in \citet{bezanson:11}.  For each galaxy, we include a variety of optical and near-IR broadband and medium band photometry to fit photometric redshifts using EAzY \citep{eazy} and calculate stellar masses using FAST \citep{kriek:09}, assuming \citet{bc:03} stellar population synthesis models and a \citet{chabrier:03} Initial Mass Function (IMF) and scaling the implied M/L ratio to the total luminosity of the best-fit S\'{e}rsic profile.  By insuring that the luminosity of the galaxy reflects the best-fit profile, and not the aperture photometry, we can more accurately correct for non-homology such that stellar mass better predicts dynamical mass (and thereby inferred velocity dispersion will better predict central velocity dispersions) \citep{taylor:10}.  We use GALFIT \citep{galfit} to measure sizes using imaging in two wavebands (UKIDDS J and K for the UDS \citep{williams:09} and ACS I and WIRDs K for the NMBS Cosmos field \citep{scoville:07,bielby:11,bezanson:11}).  Utilizing a S\'{e}rsic dependent Virial constant \citep{bertin:02}
\begin{equation}
K_v(n)=\frac{73.32}{10.465+(n-0.94)^2}+0.954
\end{equation}
and a scale factor fit to SDSS data, we calculate the inferred velocity dispersion within $r_e/8$ in both wavebands for all galaxies in the survey.  
\begin{equation}
\sigma_{inf}=\sqrt{\frac{GM_{\star}}{0.557K_v(n)r_e}}.
\label{eqn:siginf}
\end{equation}

For this analysis, we include galaxies with inferred velocity dispersions above $100\unit{km/s}$, above which the relation is calibrated locally and our samples are complete.  At all redshifts, we compare the inferred velocity dispersions calculated in each bandpass, which agree with a scatter of $\lesssim0.1\unit{dex}$ at all redshifts.  In order to evaluate the importance of including the S\'{e}rsic dependent constant and correction to total luminosity we also compare inferred velocity dispersions calculated in each bandpass with $k=5.0$ as used in \citet{cappellari:09}.  We find that in addition to increasing the scatter between the inferred dispersions in the different bandpasses by up to $0.05\unit{dex}$, eliminating corrections for non-homology introduces a net offset such that the K band inferred dispersions are larger by $\sim0.05\unit{dex}$ than those measured in the I band ACS imaging.  This offset is also apparent in the UDS, though the offset is smaller at $\sim0.01\unit{dex}$ between the J and K bands.  We note, however, that the S\'{e}rsic indices derived from ground-based data (Cosmos K band sizes and all sizes in the UDS) may have significant uncertainties (see \S5).  

Finally, to limit the effects of color gradients on our results, we interpolate the inferred velocity dispersion and sizes to a rest-frame wavelength of $6200\unit{\AA}$, the central wavelength of the r filter where the SDSS inferred dispersions are calibrated.  

\subsection{The Effects of Gas}

At the core of this analysis is the fact that the inferred velocity dispersion of a galaxy is proportional to the square root of the mass of a galaxy over its size.  At low redshift, where inferred velocity dispersion is calibrated, stellar mass is a good estimate of the baryonic mass content of a galaxy, especially for galaxies with high enough inferred dispersions to be included in this analysis.  However, at higher redshifts galaxies have higher specific SFRs \citep[e.g.][]{daddi:07,elbaz:07,noeske:07,zheng:07,damen:09}.  If we assume the local \cite{kennicutt:98} relation between SFR and gas surface density holds, this implies that the gas density and therefore the gas mass fraction for a given size should increase at higher redshifts \citep[e.g.][]{daddi:07,franx:08,nelson:12}.  Assuming that the stellar mass of galaxies included in the calibration of inferred dispersion at $z=0$ will include stars that have formed more recently, it could be important to include the gas mass in the total mass budget in estimating inferred velocity dispersions at earlier times.  

For each galaxy, we have estimates of the SFR from the SED fits.  Using these values and assuming that gas is uniformly distributed in a circular aperture of radius $r_e$, we calculate gas masses using \citet{kennicutt:98}: 
\begin{equation}
M_{gas}/M_{\odot} = \frac{(SFR/\unit{(M_{\odot} yr^{-1})})\pi (r_e/\unit{pc})^2}{(\pi(r_e/\unit{kpc})^2 (2.5\times10^{-4}))^{(1./1.4)}}
\end{equation}
We calculate inferred dispersions using $M_{baryon} \equiv\,M_{\star}+M_{gas}$ in place of $M_{\star}$.  

Although implied gas fractions increase for the catalogs as a whole with redshift, gas fractions are relatively low, even for star-forming galaxies selected in this way.  At lower redshifts, $\langle M_{gas}/M_{baryon}\rangle\sim0.05$ for galaxies above the velocity dispersion limit of the sample and the average does not evolve strongly with redshift.  This is much lower than the $35\%$ gas fractions measured by \citet{tacconi:10} in massive galaxies at $z\sim1.2$.  This discrepancy could be due to the fact that if velocity dispersion is a good predictor of quiescence \citep[e.g.][]{franx:08,wake:12b,bell:12}, then star-forming galaxies selected by velocity dispersion would, on average, have lower SFRs than those with lower velocity dispersions, and therefore lower implied gas masses; an effect that would be magnified by the fact that the velocity dispersion cut increases with redshift.  This is consistent with the fact that many of the $z\sim1.2$ galaxies in the \citet{tacconi:10} sample are consistent with being large, rotating disks with low velocity dispersions as opposed to those at higher redshifts, which might be more turbulent.  Although we are not directly measuring velocity dispersions, such massive star-forming galaxies have very large radii ($r_e\sim5\unit{kpc}$) and low S\'{e}rsic indices ($n<1$) \citep{forster:11}. Even with $M_{\star}=3\times10^{10}\,M_{\odot}$ \citep{tacconi:10}, this corresponds to $\sigma_{inf}\sim80\unit{km/s}$, below the limits of this study.    Additionally, this discrepancy could be due underestimates of the star-formation rates calculated by stellar population synthesis modeling alone due to underestimation of dust attenuation, and the effect could be slightly stronger using SFRs based on MIPS fluxes.  We note that some galaxies with high implied gas fractions are in our $\sigma_{inf}>100\unit{km/s}$ sample: in the highest redshift bin the most highly star-forming $\sim4\%$ of ($\sigma_{inf}>100\unit{km/s}$) galaxies in the Cosmos field and $\sim6\%$ of galaxies in the UDS have average implied gas fractions equal to $\sim35\%$.  We find that including the gas mass has very little effect on our final number density distributions (see \S3) with no changes to the distribution of quiescent galaxies.  At $z\gtrsim0.6$, there is a slight shift in the VDF for star-forming galaxies - at most a $\sim10\%$ effect at $\sigma\sim200\unit{km/s}$, however this shift is within the error bars.  Given the uncertainties in the SFRs, we neglect the effects of cold gas on the inferred velocity dispersions for the rest of this analysis.  For completeness, we include the VDF for star-forming galaxies with and without the inclusion of gas as Figure \ref{fig:VDF_gas} in Appendix D.  
\subsection{Surface Densities}

In addition to velocity dispersions, we calculate surface densities:
\begin{equation}
\Sigma = \frac{M_{\star}}{\pi r_e^2}
\end{equation}
for all galaxies.  This measure of galaxy compactness exhibits similar, although slightly weaker, correlations with galaxy color \citep[e.g.][]{franx:08, wake:12b} and dark matter properties \citep{wake:12a} as velocity dispersion.  Therefore we will also express our results in terms of surface density in \S3.3 to provide an important check on the results of this study.  Inferred velocity dispersions, at least those $\geq100\unit{km/s}$, are well-calibrated for galaxies of all types in the SDSS.  While there have been efforts to measure galaxy dynamics and determine the accuracy of inferred dispersion, at high redshifts \citep[e.g.][]{treu:05,wel:05,cappellari:09,dokkumnature:09,newman:10,onodera:10,bezanson:11,martinezmanso:11,sande:11} these efforts have focused on quiescent galaxies with high inferred velocity dispersions.  We emphasize that we expect the results using surface density ($\propto M/R^2$) to be similar to those derived using inferred velocity dispersion ($\propto M/R$), however, it is a more direct observational diagnostic and requires no calibration and is therefore not subject to uncertainties in evolution of constants unlike velocity dispersion.  

\subsection{Differentiating between Star-Forming and Quiescent Galaxies}

We separate into red and dead versus blue and star-forming galaxies based on rest-frame colors.  Although locally this would translate into an early type/late type split, this looser definition allows us to separate based on the stellar populations of galaxies while including the flexibility for structural evolution of entire populations of galaxies.

Using EAzY, we calculate rest-frame U-V and V-J colors for each galaxy.  Galaxies can have red U-V colors due to older stellar populations or because they are dusty.  In the case where dust is obscuring the rest-frame UV and optical light, light is re-radiated at longer wavelengths, causing these galaxies to have redder V-J colors than truly quiescent galaxies.  We therefore create a two color plot with rest-frame U-V and V-J colors to isolate red galaxies which are no longer forming stars from star-forming galaxies, using the technique described by \citet{williams:09}.  This color separation has been shown to agree well with other star formation indicators, such as MIPS $24\mu\,m$ detection \citep[see e.g.][]{williams:09,wuyts:09,bell:12} The thresholds drawn in the color-color plots to isolate the red sequence from star-forming galaxies evolve with time as the stellar populations of dead galaxies age.  We adopt the relations determined for the NMBS fields in \citet{whitaker:11} in two redshift bins ($0.3<z<0.5$ and $0.5<z< 1.5$) and show the distributions in each field and redshift range in Figure \ref{fig:UVJ}, with median random errors indicated in the lower right corner.  The overall galaxy distribution in a given redshift range is indicated by the greyscale density, including galaxies below the velocity dispersion threshold, which we do not include in our analysis.  Colored points indicate galaxies with $\sigma_{inf}\geq100\unit{km/s}$: star-forming galaxies are identified with blue points and quiescent with red.  In each plot, the red sequence of quiescent galaxies is characterized by a diagonal sequence in the upper left region of color-color space.  We note that although we define a bimodal population, there is a continuum of galaxy ages and ongoing SFRs.  Based on the correlation between velocity dispersion and galaxy type, we expect that even the star forming galaxies in the sample may not be strongly forming new stars due to the cut in velocity dispersion, as suggested by the minimal impact of including gas mass (as discussed in \S2.1).

We evaluate the effect of defining quiescence using these color cuts by repeating the analysis using cuts on star-formation rates derived from IR and UV luminosities in the Cosmos field, where MIPS 24 $\mu m$ data is included in the NMBS catalogs \citep{whitaker:11,whitaker:12b}.  We make a quiescent distinction cut for star formation rates greater than twice the standard deviation below the redshift-dependent star-forming main sequence as defined by \citet{whitaker:12b}.  Measured velocity dispersion functions for star-forming and quiescent galaxies using this method agree extremely well with those based on color cuts, after applying a correction for the $24 \unit{\mu m}$ incompleteness as a function of $\sigma$.  The color derived quiescent population is less than $10\%$ contaminated by MIPS detected sources, and down to $1-2\%$ at the highest and lowest redshifts, corresponding to deviations from the observed VDF by an average of $\lesssim0.1\unit{dex}$ at all velocity dispersions.  Because of the additional complexity of understanding incompleteness in MIPS fluxes, we focus all further analysis on color-based quiescence as it is complete for all galaxies in the surveys.  We note that this test is performed in the NMBS Cosmos field, which may have more accurate rest-frame colors due to the improved sampling of the NIR medium band filters.  There is a hint in Figure \ref{fig:UVJ} that the two populations of galaxies are less distinct at higher redshifts in the UDS field than in the NMBS field and therefore the quiescent sample may be slightly more contaminated.  We emphasize that although there are field-to-field offsets in the overall VDF, there does not appear to be a systematic trend in the relative fractions of star-forming and quiescent galaxies as a function of dispersion.

\begin{figure}[!t]
  \centering
  \epsscale{1.}
  \plotone{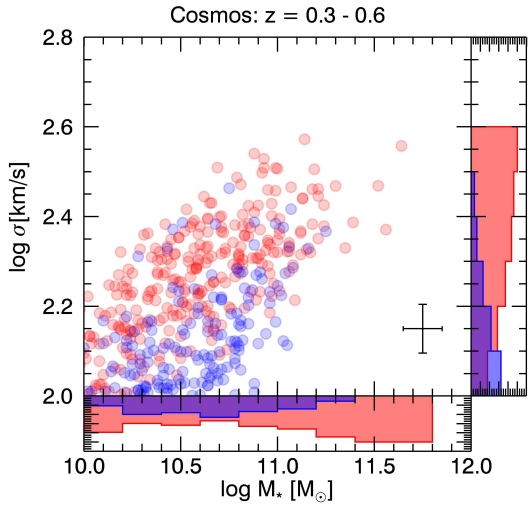}
    \caption{Inferred dispersion versus stellar mass in Figure \ref{fig:msigma} in the COSMOS field for $0.3<z\leq0.6$, red points represent quiescent galaxies and blue points star-forming galaxies. Histograms on the bottom and right side indicate the relative contributions of the two populations.  It is clear from these histograms that velocity dispersion of a galaxy is a better predictor of quiescence in galaxies than stellar mass.  Approximate errors are indicated in the lower right corner.}
  \label{fig:msigma}
\end{figure}

In order to emphasize the importance of using velocity dispersion as opposed to stellar mass as an indicator of stellar properties, we plot the inferred dispersion versus stellar mass in Figure \ref{fig:msigma} in the COSMOS field for $0.3<z\leq0.6$ where the NEWFIRM medium band near-IR filters begin to contribute to the excellent photometric redshifts.  It is clear from the histogram on the right that velocity dispersion produces a stronger separation between quiescent and star-forming galaxies than stellar mass (bottom histograms). 

\section{Number Densities}

\subsection{Velocity Dispersion Function for Star-Forming and Quiescent Galaxies}

\begin{figure*}[!ht]
  \centering
  \begin{tabular}{cc}
	\includegraphics[scale=0.15]{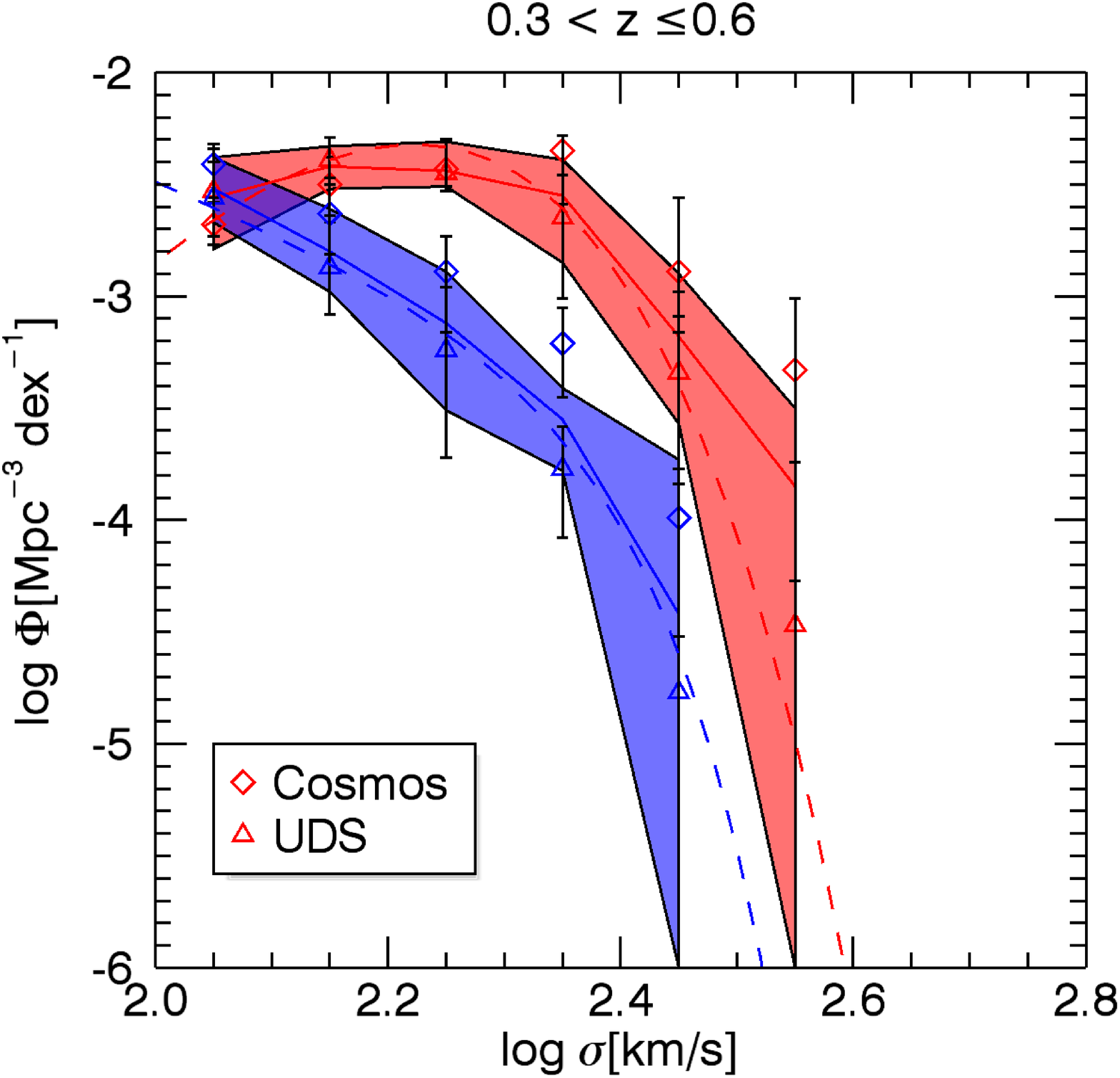} &
	\includegraphics[scale=0.15]{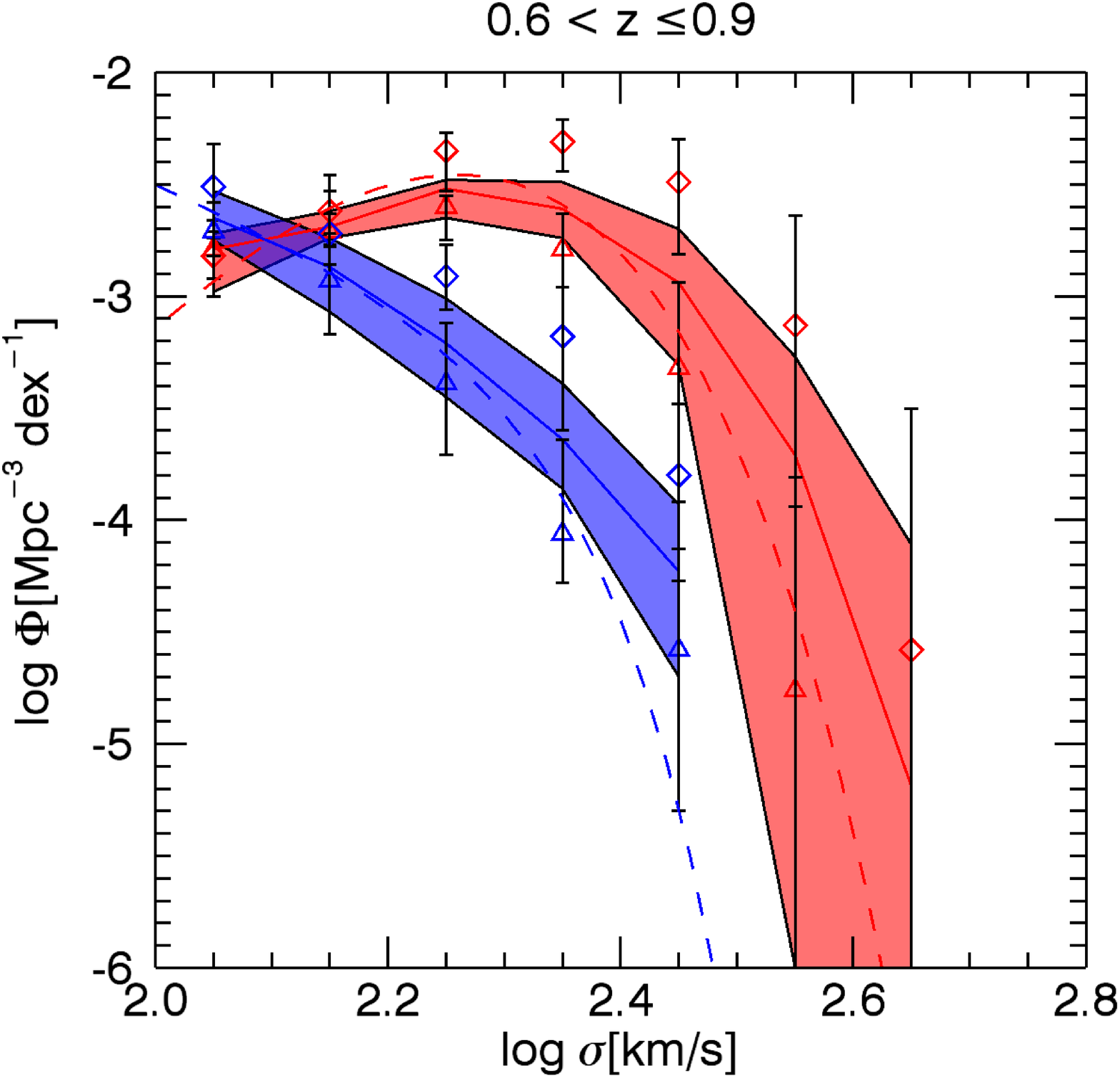} \\
	\includegraphics[scale=0.15]{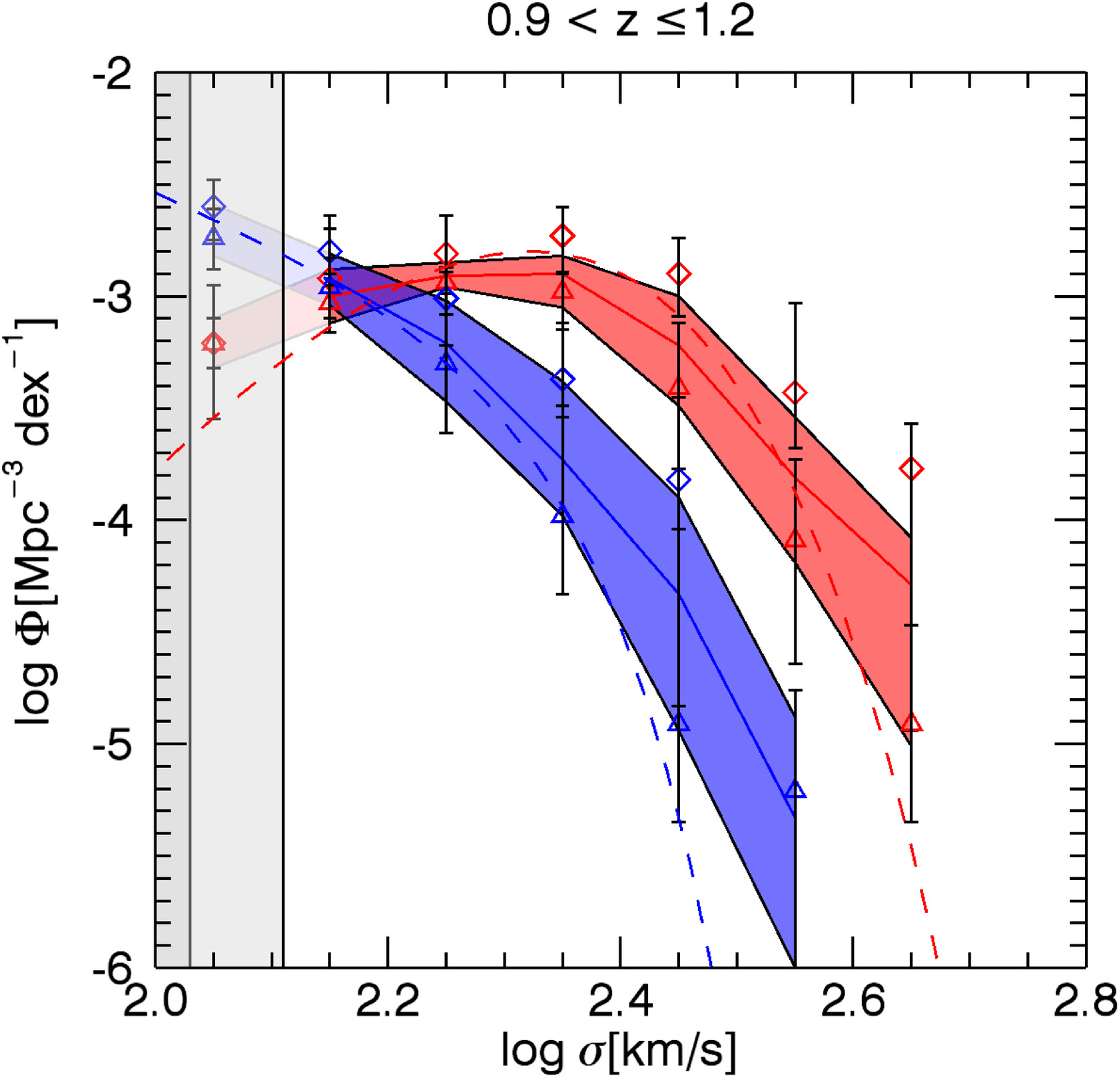} &
	\includegraphics[scale=0.15]{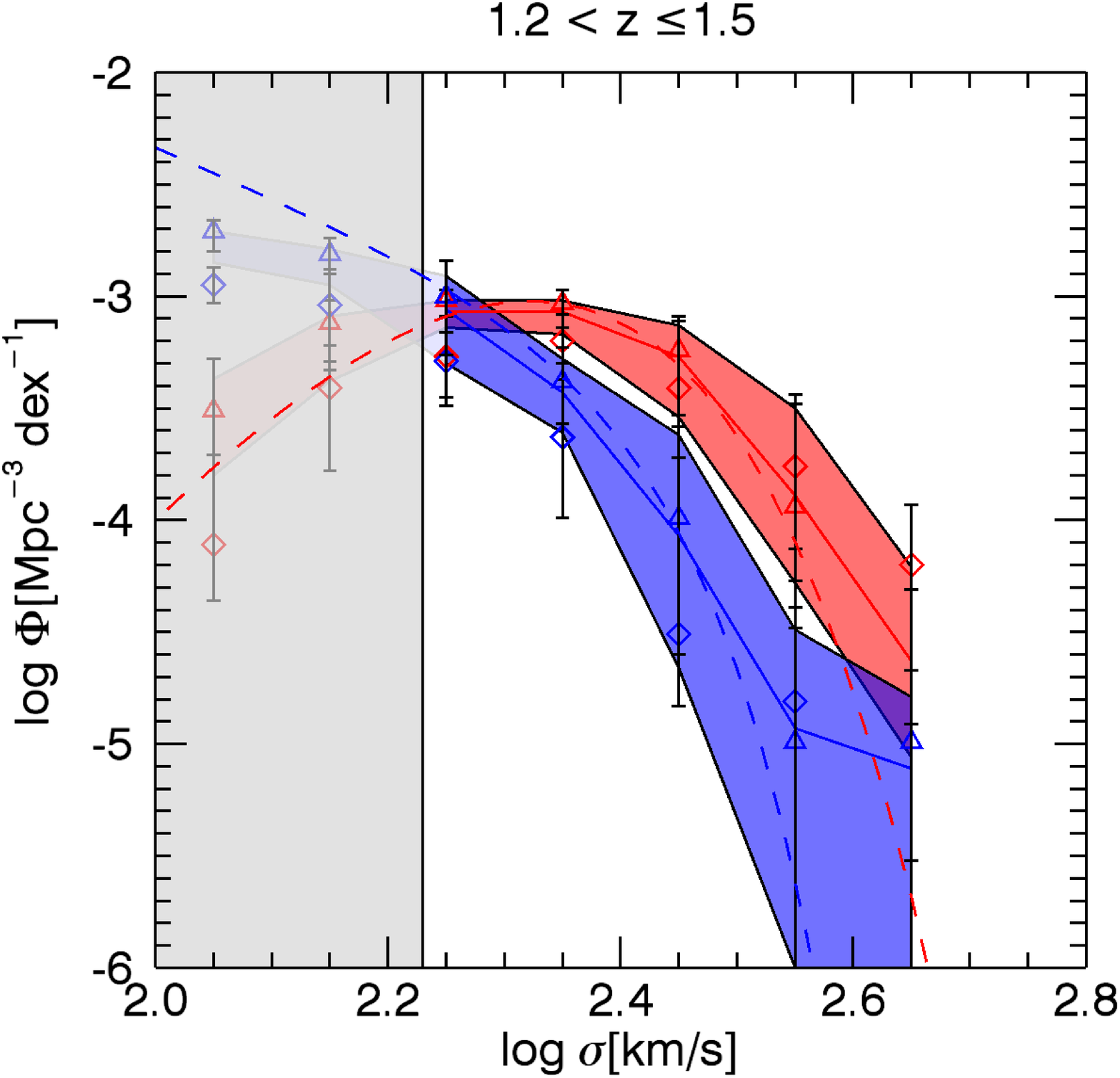} \\	
   \end{tabular}
    \caption{The VDF for quiescent (red) and star-forming (blue) galaxies as a function of redshift.  Data points represent the measured VDF in Cosmos (diamonds) and UDS (triangles) fields individually.  Large, light-colored regions show the combined VDF for the two galaxy populations at each redshift.  Estimated incompleteness limits at a given redshift range for each photometric field are included as grey shaded regions.  Here we see at all redshifts that the VDF for quiescent galaxies turns over at an intermediate velocity dispersion and falls off steeply at high inferred velocity dispersions.  In contrast, the number density as a function of velocity dispersion is roughly flat for star-forming galaxies in all redshift bins.}
  \label{fig:VDF}
 \end{figure*}

\begin{figure*}[!ht]
  \centering
  \begin{tabular}{cc}
	\includegraphics[scale=0.15]{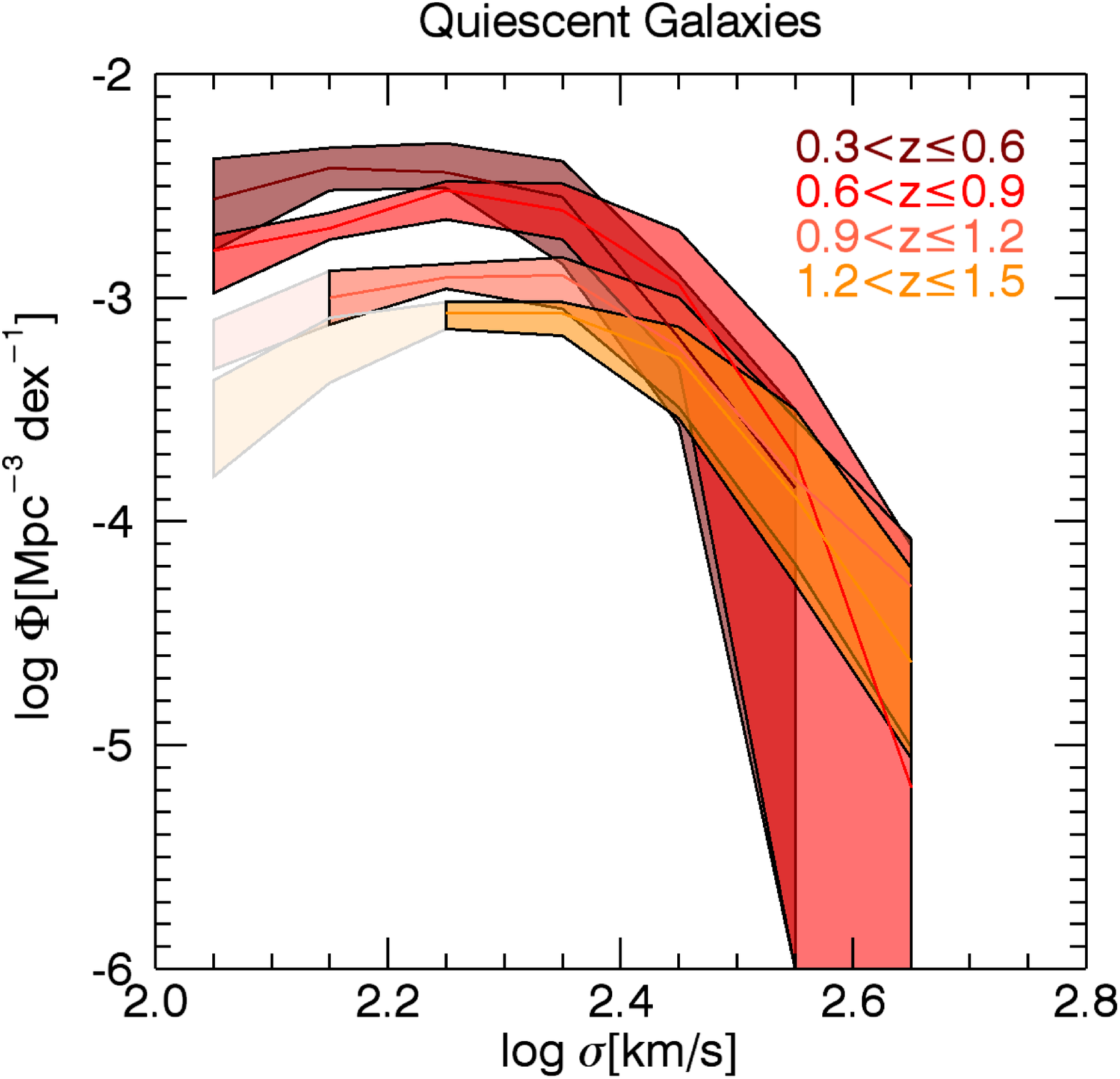} &
	\includegraphics[scale=0.15]{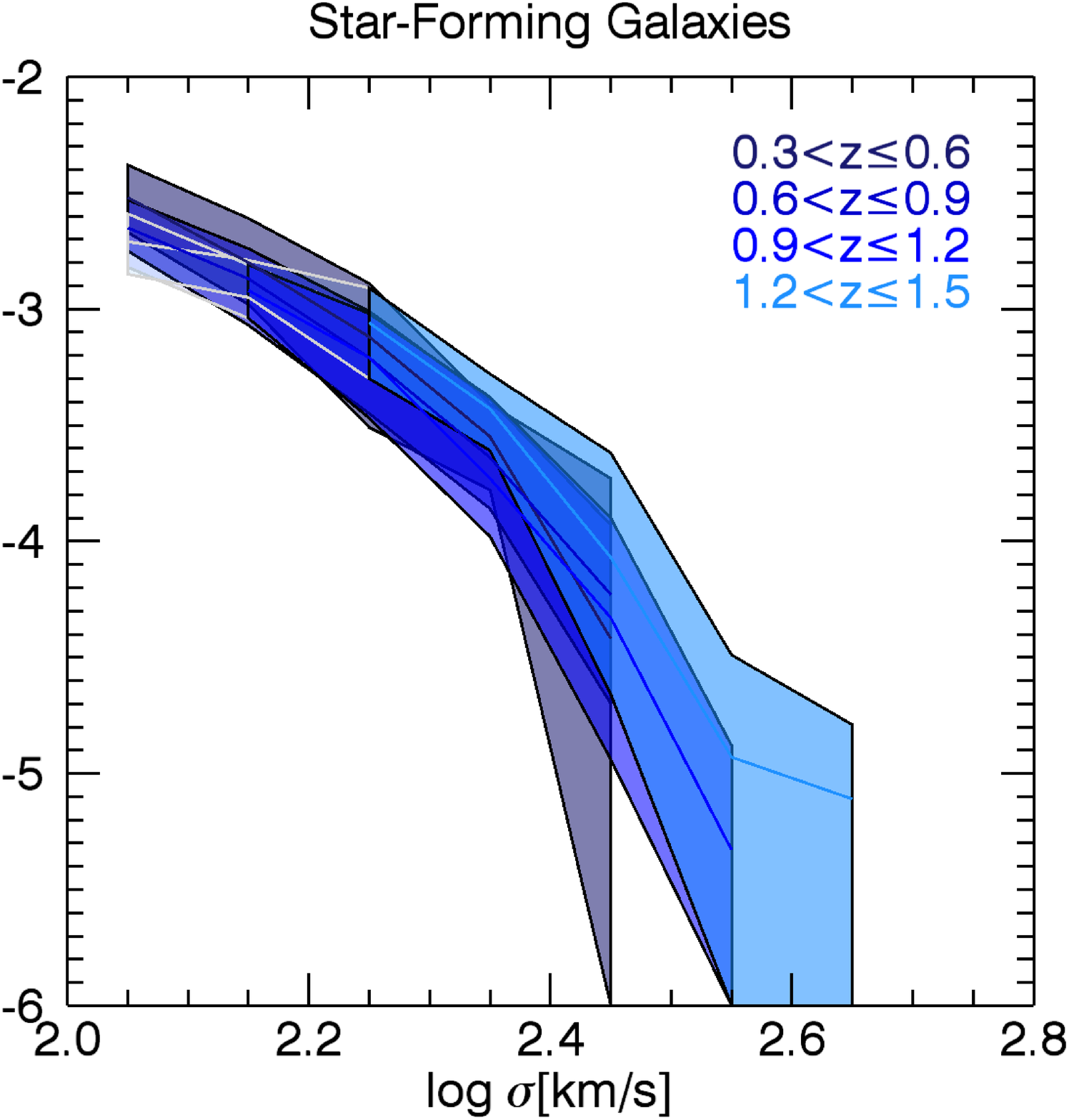} \\
	\includegraphics[scale=0.15]{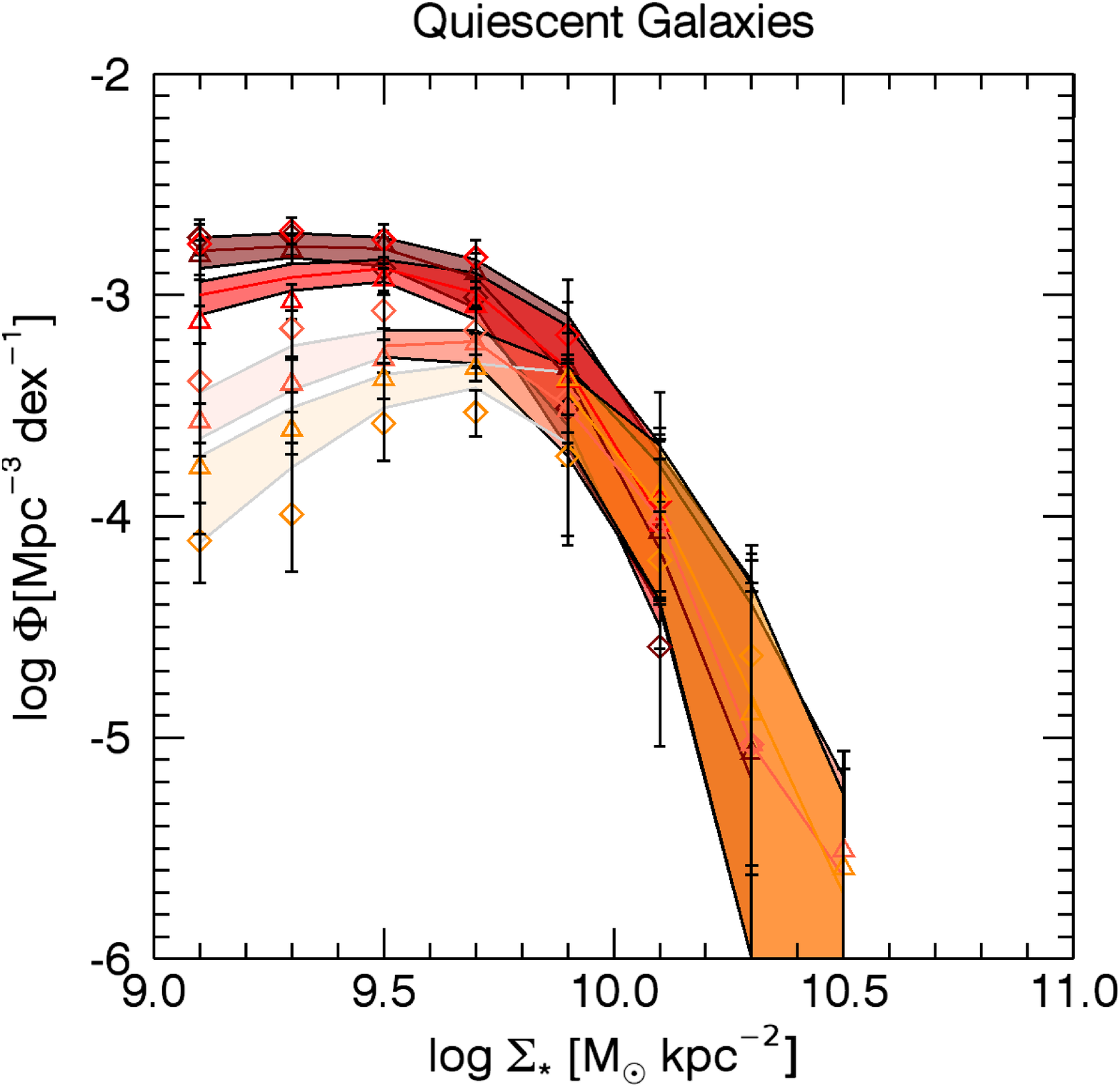} &
	\includegraphics[scale=0.15]{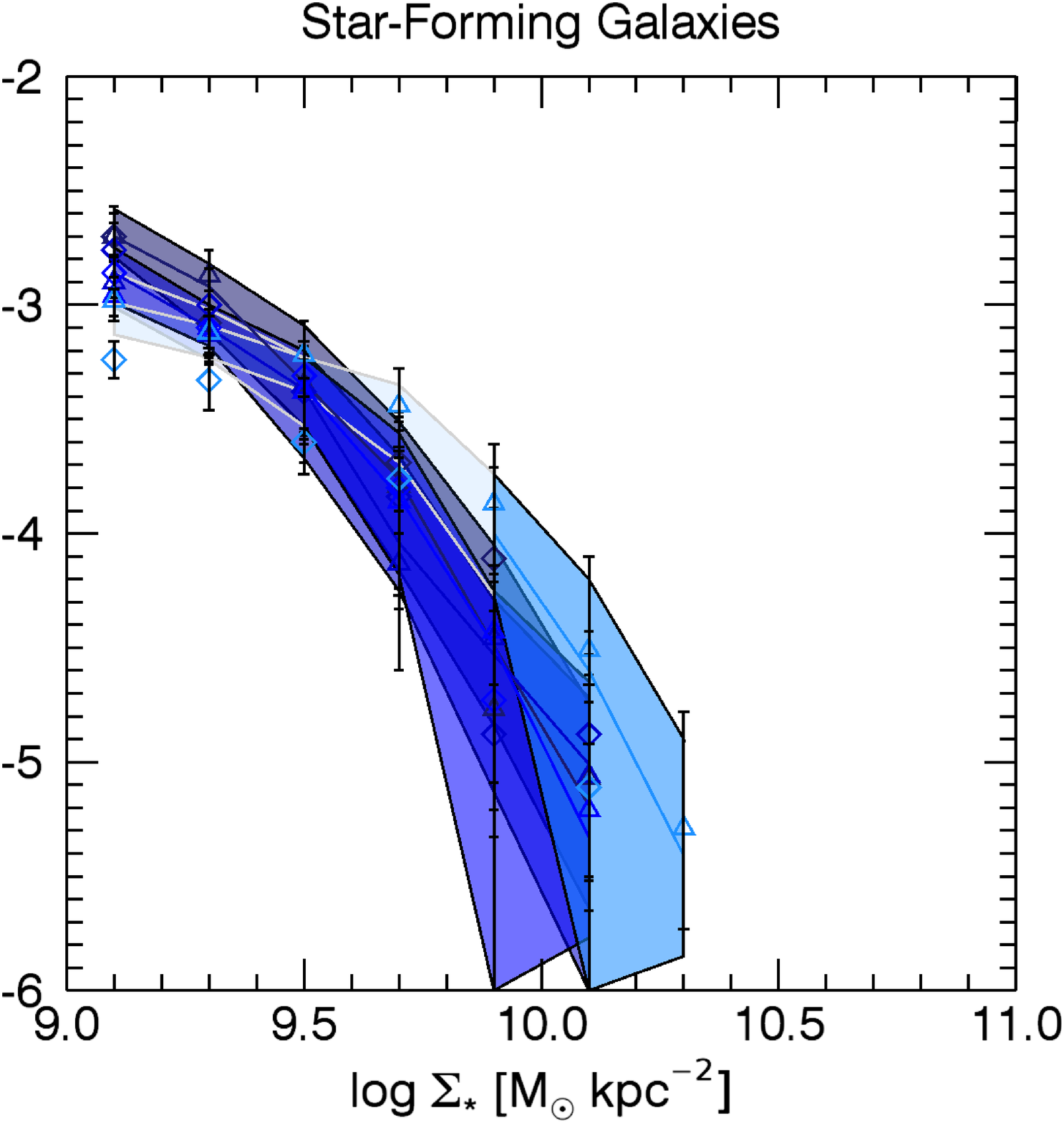} \\
  \end{tabular}
    \caption{\emph{Top Row:} VDFs for Quiescent (left panel) and Star-Forming (right panel) Galaxies at all redshifts.  We note that while the quiescent VDF evolves strongly, building up low dispersion galaxies with time, the shape of the VDF for star-forming galaxies remains mostly constant. \emph{Bottom Row:} Surface density functions for quiescent and star-forming galaxies with Poisson error bars.  The differential evolution of number density for quiescent galaxies is also apparent as a function of surface density, as is the roughly constant shape for star-forming galaxies, at least out to $z\sim1.2$.}
  \label{fig:VDF_MR2}
 \end{figure*}

Once we have calculated inferred dispersions and categorized every galaxy as quiescent or star-forming, we calculate the number density as a function of velocity dispersion for each field and population separately, included as open symbols (diamonds for the COSMOS field and triangles for the UDS) in Figure \ref{fig:VDF}. There is some field-to-field variation, however the qualitative distribution of galaxies is the same for the COSMOS and UDS fields. The volume-weighted average VDFs for quiescent and star-forming galaxies are shown in red and blue polygons.  Error bars on all points reflect Poisson errors in addition to systematic errors due to cosmic variance, photometric redshifts and stellar masess (see \S3.2).  Incompleteness in $\sigma_{inf}$ is dominated by galaxies that are too faint to measure reliable sizes and are therefore not included in the size catalogs ($K\geq22.4$ in the UDS and $K\geq22.0$ in the COSMOS field). We define the dispersion completeness limit (grey regions in Figure \ref{fig:VDF}) for each field and redshift range as the dispersion at which the $95\%$ completeness plus $0.06\unit{dex}$ scatter about a linear fit to the K magnitude - inferred dispersion relation reaches the magnitude limit of the size catalogs (see Appendix A).  

At every redshift, the shape of the quiescent VDF is very similar to the modified Schechter function that is fit to the VDF of elliptical galaxies in the SDSS \citep[e.g.][]{sheth:03,mitchell:05,choi:07,shankar:09,bernardi:10}.  On the other hand, the VDF for star-forming galaxies is primarily flat in log-log space and is much steeper than the quiescent VDF except at the highest dispersions.  At all redshifts, star forming galaxies are much less common than quiescent galaxies at velocity dispersions greater than $\sim150\unit{km/s}$.  

In this analysis we construct the VDFs from inferred, not measured velocity dispersions. Therefore, the underlying distribution of intrinsic velocity dispersions is broadened by scatter in the $\sigma_0$ vs. $\sigma_{inf}$ relation, which becomes especially dominant at the steep, high dispersion end of the VDF.  This is essentially a form of Eddington bias whereby high $\sigma_{inf}$ galaxies are likely to have intrinsically lower $\sigma_0$ values that have been scattered upwards.
We account for this by simulating an underlying distribution following a modified Schechter function for the quiescent galaxies and a double power law for the star-forming galaxies, scattering by the observed scatter of $0.06\unit{dex}$ from the SDSS \citep[see][]{bezanson:11}, and then fitting the scattered distribution to the observed VDFs.  The underlying distributions are included as blue (star-forming) and red (quiescent) dashed lines in Figure \ref{fig:VDF}.  This scatter should be interpreted as a minimal scatter, as additional sources of uncertainty such as redshift errors and systematic errors in stellar masses might increase with redshift.  In every case, the unscattered VDF agrees with the scattered VDF at the low dispersion end.  However, since the number density of very high dispersion objects is low, adding scatter to the velocity dispersion in the model broadens the intrinsic VDF.  Given an increased scatter above the nominal $0.06\unit{dex}$ from SDSS the broadening of the intrinsic VDF would be stronger, implying an even steeper drop in the number density of galaxies with high velocity dispersions.  This effect is relevant when interpreting the measured properties of small samples of high redshift galaxies selected on inferred velocity dispersion.

To highlight the evolution of each population with time, we plot the VDF for each galaxy type at all redshifts in the top row of Figure \ref{fig:VDF_MR2}.  \citet{bezanson:11} demonstrated that the shape of the overall VDF has evolved strongly at the low velocity dispersion end, but that the overall number density of galaxies at high inferred velocity dispersions evolved very little since $z\sim1.5$. By separating into quiescent and star-forming galaxies, we can gain insight into how this occurred. In general, the shape of the quiescent galaxy VDF remains similar in all redshift bins, but the low velocity dispersion end builds up with time. The number of star-forming galaxies at fixed dispersion is remarkably constant at all redshifts, perhaps with a slight increase at high dispersions in the highest redshift bin.  Another key feature is that there is very little evolution in the number of high velocity dispersion galaxies of either type.  The strong evolution of quiescent galaxies and roughly stable number density of star forming galaxies has also been studied as a function of stellar mass \citep[e.g.][]{brammer:11}.
 
 \subsection{Uncertainties in the Velocity Dispersion Function}

There are clear discrepancies between the number densities of galaxies in the two fields, often much larger than the associated Poisson errors.  We will here try to evaluate the impact of these sources of error on the overall VDFs and include all error estimates along with the measured velocity dispersion functions in Table \ref{tbl:VDF}.  We take the following sources of systematic error into account: cosmic variance estimated using \emph{QUICKCV} \citep{moster:11}, photometric redshift uncertainties and errors in stellar mass estimates.  Because the UDS field relies only on broadband photometry, we also investigate the possible influence of random errors on the photometric redshifts for the field.  

First, we we adopt the methodology of \citet{somerville:04, moster:11} to estimate the effects of cosmic variance on the measured VDFs from individual fields.  We use \emph{QUICKCV} \citep{moster:11} to estimate the dark matter variance based on the volume probed by each redshift bin and survey and estimate the bias at each velocity dispersion based on the bias-number density relation in \citet{somerville:04}.  In each case, the Poisson errors are larger than those expected by cosmic variance alone.  However, we note that the difference between the two fields is often much greater than $1-\sigma$ cosmic variance and Poisson statistics would predict; the number densities of galaxies in the Cosmos field are higher than those in the UDS field in all but the highest redshift bin.  The NMBS Cosmos field appears to be an overdense field, especially relative to the larger UDS field.  We include the redshift distribution of galaxies in the surveys (above the minimum $\sigma_{inf} > 100\unit{km/s}$ threshold to be included in this study) in Figure \ref{fig:redshifts}.  In this figure, the large overdensities of galaxies at $z\sim0.72$ and $z\sim0.88$ in the Cosmos field are immediately apparent relative to the panel to the right showing the number density of galaxies in the UDS field in the same redshift range.  These peaks fall in the same VDF redshift bin and can therefore explain the maximum of field to field variation observed at those redshifts.

The two surveys included in this analysis are both extremely deep, well-calibrated samples of galaxies, however the NMBS Cosmos survey includes photometry in near-IR medium band filters that are explicitly designed to sample the Balmer/$4000\unit{\AA}$ break for galaxies at $z\gtrsim1$, allowing for much better sampling of the SED and more accurate photometric redshifts.  Therefore, while the UDS field should reach $\sim3\%$ accuracy in $\frac{\Delta z}{1+z}$ \citep{williams:09}, the NMBS field reaches an impressive $\sim1\%-2\%$ accuracy.  Photometric redshift uncertainties effect both the redshift bins in which galaxies fall, but also fold into their stellar masses and physical sizes.  To determine the impact of systematic errors in redshifts on the final VDFs, we calculated the scatter in the VDF due to introducing redshift shifts to each of the fields, $\pm1\%$ for Cosmos and $\pm3\%$ in the UDS.  We note that random errors in photometric redshifts would introduce a much more subtle effect as photometric redshifts would scatter in either direction and would produce smaller error bars than this extreme estimate.  The effect of systematically shifting redshifts on the dispersion function is generally small, $\lesssim0.1\unit{dex}$, comparable to the Poisson errors on each field.  This is not the case, for example, in the Cosmos field between $0.6<z\leq1.2$, where the errors grow larger ($\pm\sim0.2-0.3$ or more).  This can be explained by the tail of the $z\sim0.9$ overdensity shifting between the two redshift bins.  

Systematic shifts in photometric redshifts will have a stronger impact on the overall VDF than random errors, however if the $3\%$ accuracy of photometric redshifts is an underestimate in the UDS, these could become more important.  To test this effect, we adopt a high estimate for random photo-z uncertainty, introducing a gaussian scatter the photometric redshifts with a width of $8\%$ in $\Delta\,z/(1+z)$, and refit stellar masses and calculate the final VDF.  We verify that the effect of this test is quite minimal, introducing a mean offset of $0.04\unit{dex}$ (with an rms of $0.065\unit{dex}$) for the quiescent VDF and $0.05\unit{dex}$ (rms$=0.169\unit{dex}$) in the VDF for star-forming galaxies.  In every case the offset is less than the total uncertainty on the VDF for the field as quoted in Table 1.  We conclude that the inclusion of systematic errors in photo-zs provides a sufficient description for redshift uncertainties in the UDS.  We note that this effect could indeed be increasing scatter in measured quantities and further broadening the high dispersion end of the VDF, even though we do not see evidence that the VDF in the UDS is shallower at the high velocity dispersion end than in the NMBS Cosmos field.

Stellar masses are derived from SED fitting in both surveys, a method that is riddled with systematic sources of uncertainty.  We adopt a $0.1\unit{dex}$ error in the stellar mass to allow for systematic uncertainties due to assumptions about IMF, star-formation histories and stellar population synthesis templates.   We recalculate the velocity dispersions and velocity dispersion functions for $\log M_{\star} \pm 0.1 \unit{dex}$.  This has a very direct effect on the number densities of galaxies as they shift horizontally as galaxies shift to lower, and higher, bins in $\log \sigma_{inf}$, more strongly broadening the high dispersion end of the VDFs.  These errors are nearly uniformly larger than other errors, especially for the VDF of star-forming galaxies, although the effect is slightly minimized near the peak of the quiescent VDF at $\sigma_{inf}\sim2.3$, just due to the shape of the distribution.

Altogether, the error estimates would add in quadrature and the individual fields are combined using a volume-weighted average.  At most redshifts and velocity dispersions systematic errors in stellar mass dominate the total error budget on the dispersion function.  We include the total errors for every field as error bars on the individual points in Figure \ref{fig:VDF}.  Shaded regions in Figures \ref{fig:VDF}, top panels of \ref{fig:VDF_MR2}, and \ref{fig:simevolve} reflect the volume-weighted VDFs with the combined total errors of the complete sample.  By including these various sources of systematic error, the field to field variation of the VDF has become consistent with the total errorbars, which suggests that while we cannot discriminate amongst these sources of error, at least some of them must be important for proper interpretation of the VDF as measured with this data.

 \subsection{Quiescent Fraction as a Function of Velocity Dispersion}

\begin{figure*}[!t]
  \centering
  \epsscale{1.}
  \plottwo{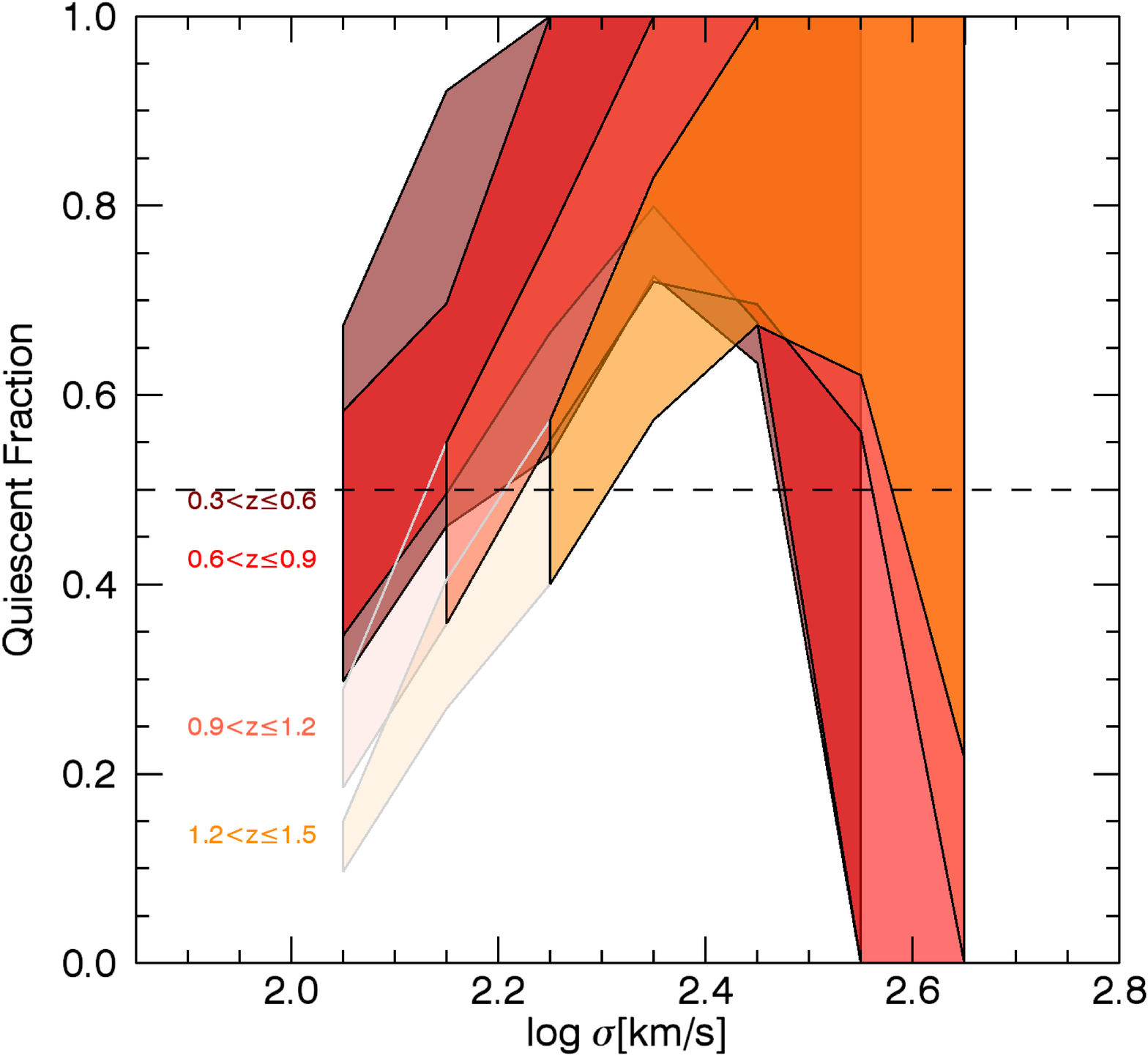}{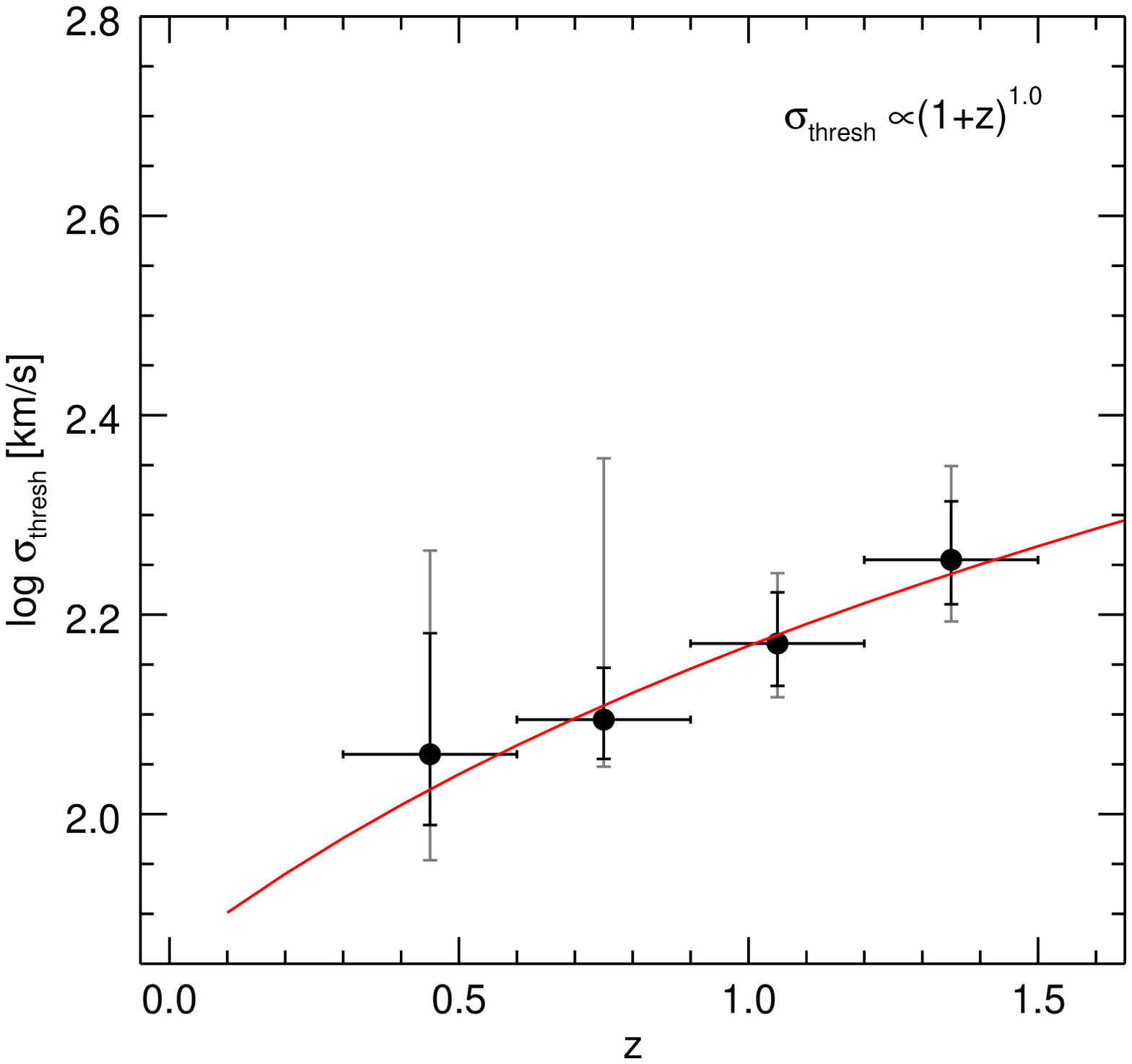}
    \caption{\emph{Left Panel:} Fraction of quiescent galaxies as a function of inferred velocity dispersion.  The dashed black line indicates the $50\%$ threshold between quiescent or star-forming dominance.  At all redshifts, the quiescent galaxies dominate at high velocity dispersion, however the crossover at which quiescent galaxies become more prevalent than star forming galaxies evolves to higher velocity dispersions at higher redshift.  \emph{Right Panel:} Evolution of the threshold velocity dispersion between quiescent and star-forming galaxies as a function of redshift.  This cross-over velocity dispersion decreases with time to the power $\sigma_{thresh}\propto(1+z)^{1.0}$, as denoted by the red line.}
  \label{fig:qfrac}
\end{figure*}

Another way of quantifying the build up of low dispersion quiescent galaxies is to plot the quiescent fraction of galaxies as a function of inferred velocity dispersion, shown in the left panel of Figure \ref{fig:qfrac}.  Unsurprisingly, the fraction of quiescent galaxies increases with velocity dispersion in each redshift bin.  However, an important result of the buildup of low-dispersion quiescent galaxies and the constant star-forming VDF is that the dispersion above which quiescent galaxies dominate becomes lower with time: from $\sim160\unit{km/s}$ in the $1.2<z\leq1.5$ down to the lowest velocity dispersions included in this analysis of $\sim120\unit{km/s}$ by $0.3<z\leq0.6$. This result supports a picture of downsizing of galaxies, with galaxies turning off at lower dispersions, i.e. larger radius and/or smaller mass, and building up the red sequence at later times.  We note that this trend is most significant when the systematic errors in stellar mass cancel out in calculating quiescent fractions; black error bars reflect Poisson, cosmic variance and systematic redshift errors only, grey error bars in Figure \ref{fig:qfrac}b include all errors discussed in \S3.2.  This result is consistent with the observational result that quiescent galaxies with low dispersions have systematically younger ages in the SDSS \citep[e.g.][]{haiman:07,shankar:09}.  The right panel of Figure \ref{fig:qfrac} shows the threshold velocity dispersion as a function of redshift and a fit such that $\sigma_{thresh} \propto (1+z)^{1.0}$.  

\subsection{Surface Density Function for Star-Forming and Quiescent Galaxies}

In the local Universe, inferred velocity dispersion is well determined by stellar mass and structural parameters of galaxies - regardless of morphology or star-formation rates. However, all measurements of the velocity dispersions of galaxies at high redshifts are made for quiescent galaxies and inferred velocity dispersion is essentially uncalibrated for star-forming galaxies at $z>0$.  We therefore investigate whether the number density as a function of surface density, calculated in \S2.4, which may be a more observationally motivated quantity for star-forming galaxies, exhibits similar evolutionary behavior as the inferred VDF.  We present the ``surface density function" (see also the ``compactness function" in \citet{newman:12}) for quiescent and star-forming galaxies at all redshift bins in the bottom row of Figure \ref{fig:VDF_MR2}, emphasizing that the overall shapes and evolution of these distributions are quite similar to the VDFs (shown in the top row).  Symbols and colors are the same as for the VDFs and completeness limits are derived as described in \S2.1 by using the scatter about the K magnitude -- surface density relation.

We note that while the surface density function is an important test of these results, the S\'ersic dependence of inferred velocity dispersion could make the VDF more relevant for comparison with simulations because it can account for the structural evolution that has occurred in galaxies in this redshift range \cite[e.g.][]{wel:11,buitrago:11}.  Furthermore, while there may not be physical constraints on the rate at which galaxy surface densities can evolve with time, the mechanisms that can increase or decrease the motion of stars in a galaxy are more constrained.  

\section{A Simple Explanation for the Coevolution of the Quiescent and Star-Forming VDFs}

\subsection{Model Overview}

\begin{figure}[!t]
  \centering
	\includegraphics[scale=0.4]{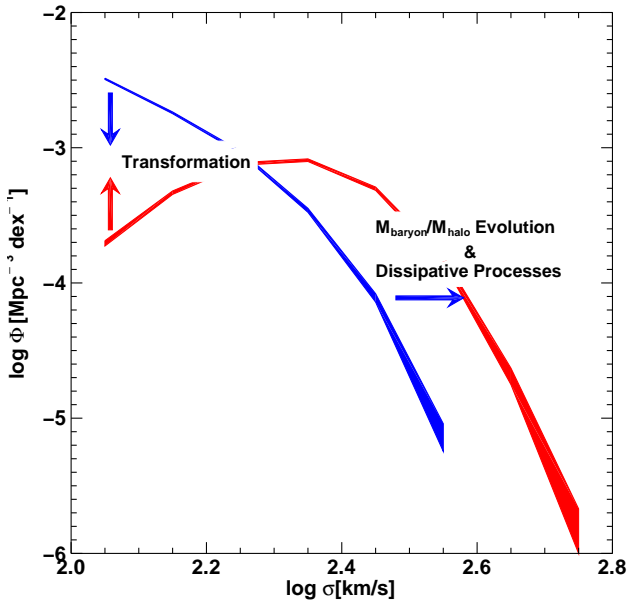}
    \caption{Simulated VDFs from the highest redshift bin with arrows illustrating the effects of transformation and dissipative processes on the shape of the VDF.}
  \label{fig:simevolve_cartoon}
 \end{figure}

As we discuss in \S1, the velocity dispersion of an individual galaxy can also evolve and we can make some simplistic statements about the physical processes that could be responsible for these changes. Gas dissipation would increase the velocity dispersion of a galaxy, which would be primarily important for star-forming galaxies.  Additionally, the contribution of dark matter to the dynamical mass of a galaxy could evolve with redshift, which would alter the ratio between the stellar and dynamical mass for a galaxy.  This would therefore change the galaxy's inferred velocity dispersion just due to evolution of the dark matter halo. Merging of galaxies could either increase or decrease the velocity dispersion of a galaxy, depending on the initial orbital energies of the progenitor galaxies and their relative masses and gas content.  For example, it has been suggested that minor merging could puff up galaxies, thereby decreasing the mass within a given radius and decreasing velocity dispersion \citep[e.g.][] {boylan:05,boylan:06,bezanson:09, naab:09,oser:11}.  Finally, gas outflows could decrease the velocity dispersion of a galaxy as mass is expelled from the central regions of the galaxy.  In addition to changes in the velocity dispersion of individual galaxies in these samples with time, the number density of the star-forming and quiescent galaxies can change due to transformation between the populations.

In order to demonstrate the qualitative effects of these processes on the quiescent and star-forming VDFs, we produce a simple model in which the number density of galaxies initially follows the underlying VDFs fit to the $1.2<z\leq1.5$ redshift bin and evolves forward in time (also see \citet{bell:07,walcher:08,peng:10} for similar analysis based on the evolution of the stellar mass function).  Without assigning physical mechanisms as responsible for changes in the galaxy population, we allow individual star-forming galaxies to quench and transform into quiescent.  While this simple evolution could reproduce the build-up of the quiescent VDF, it would deplete the population of star-forming galaxies and contradict the observed constancy of the VDF for star-forming galaxies.  Therefore, in order to maintain a constant distribution of star-forming galaxies, we allow individual galaxies to undergo net shifts in velocity dispersion in a redshift interval, $d\log\sigma/dt$.  The effects of these two processes on the VDFs are indicated in Figure \ref{fig:simevolve_cartoon}.  In this case, transformation moves galaxies from the blue star-forming VDF to the red quiescent VDF, decreasing the number density of the former and increasing the latter.  Net increases in velocity dispersion would shift the star-forming VDF to the right.  

\subsection{Model Fitting}

In this context, the increase in low-dispersion quiescent galaxies can simply be explained by the transformation of star-forming galaxies where they are most common - at the low dispersion end of the VDF.  We set the transformation rate equal to the increase in the observed VDF for quiescent galaxies at a given velocity dispersion and in each redshift bin as determined by the unscattered fits.  We impose a minimum transformation rate of zero for velocity dispersions at which the implied evolution is negative.  Adding only transformation to the model quickly depletes the number of star-forming galaxies, whereas the observed star-forming VDF is constant with time.  Therefore, we introduce a net change in the velocity dispersions of star-forming galaxies at a given velocity dispersion, such that the number of galaxies leaving the bin in velocity dispersion equals the number of galaxies entering from the adjacent bin.  Matching the number flux insures the constant shape of the star-forming VDF with time.  At each timestep, the velocity dispersions of individual galaxies are randomly scattered by $0.06\unit{dex}$ to reflect the scatter in measured and inferred velocity dispersions in the SDSS and the VDF is recalculated.

\begin{figure*}[!t]
  \centering
  \begin{tabular}{cc}
	\includegraphics[scale=0.35]{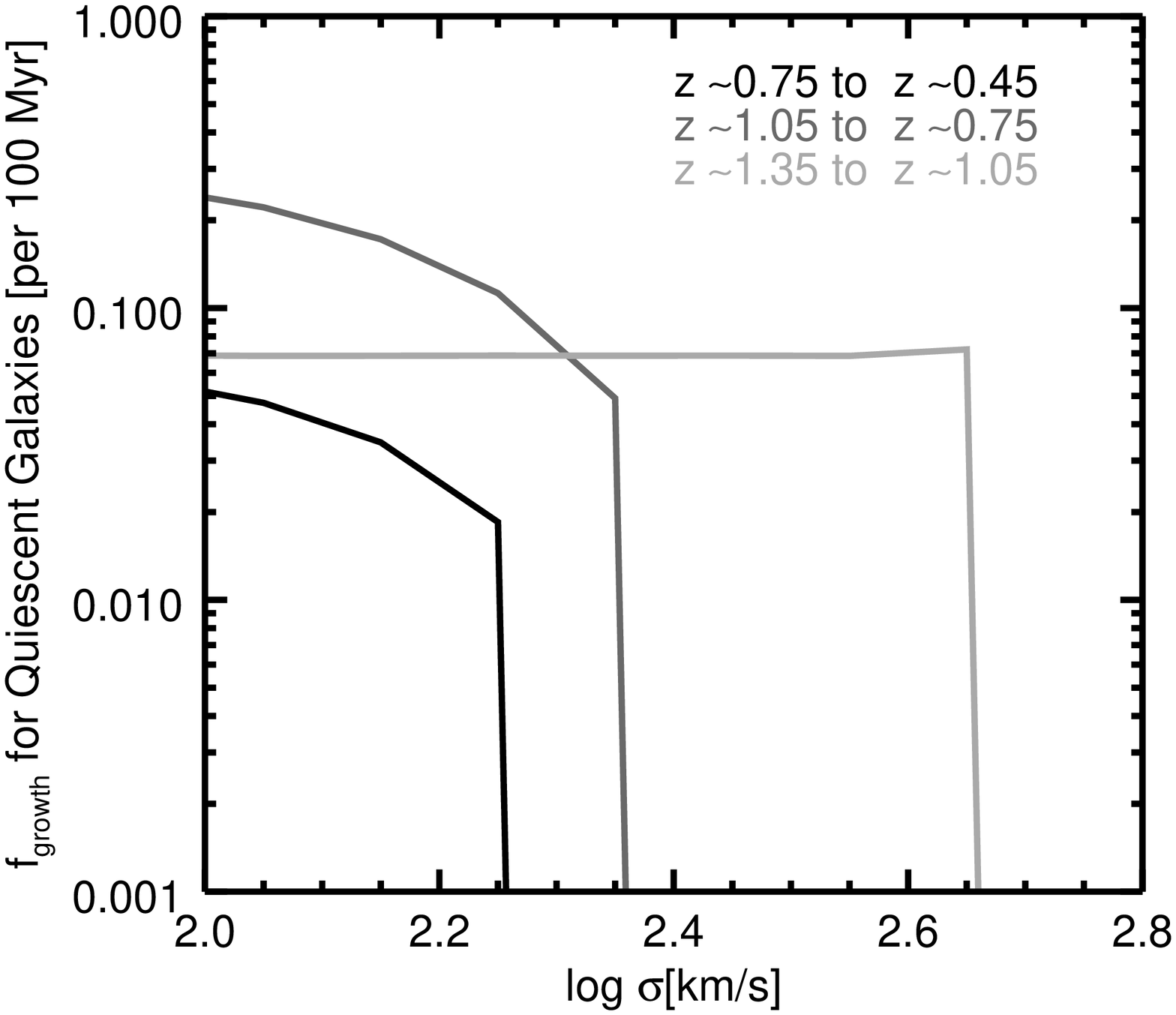} &
	\includegraphics[scale=0.35]{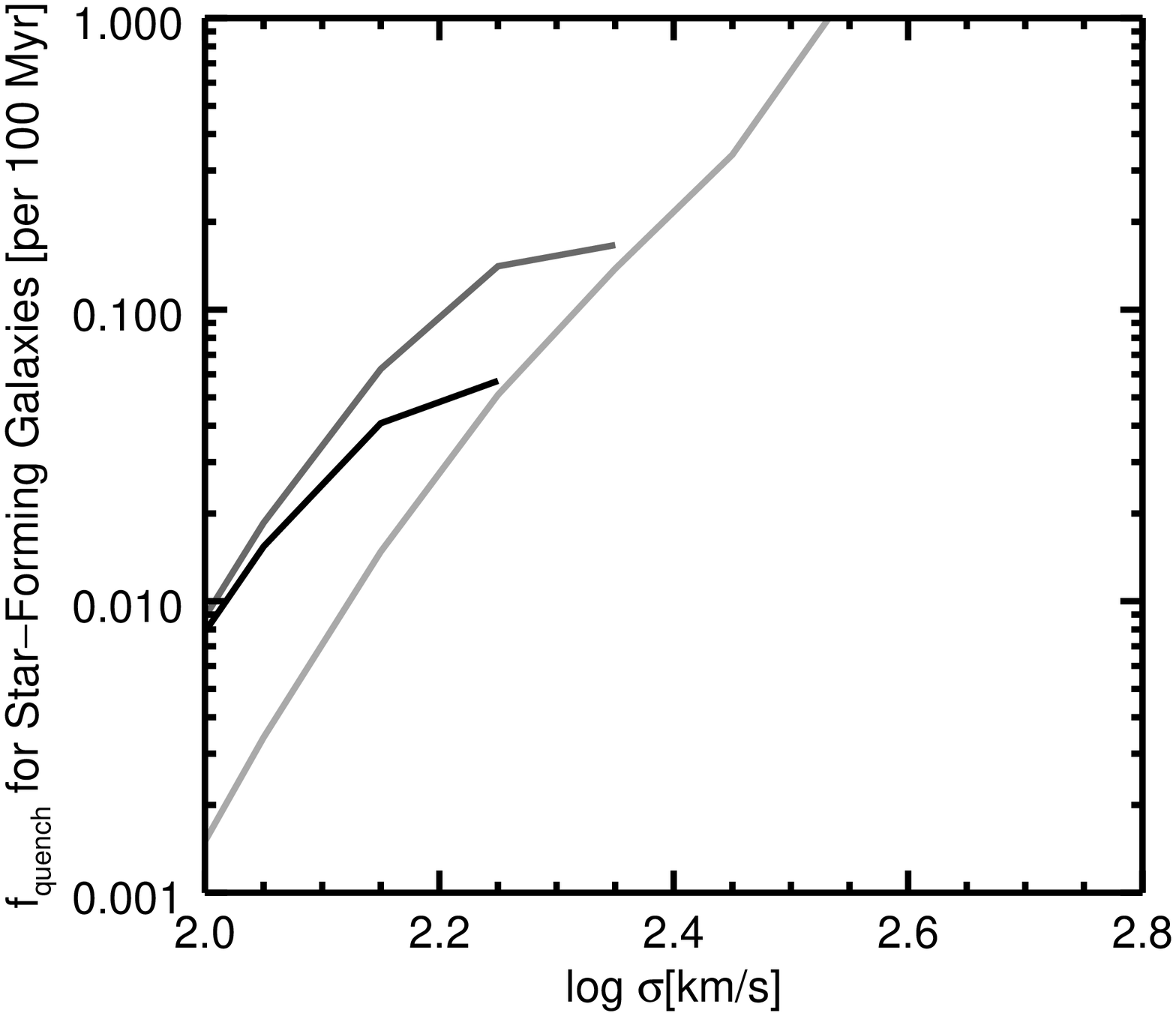} \\
	\includegraphics[scale=0.35]{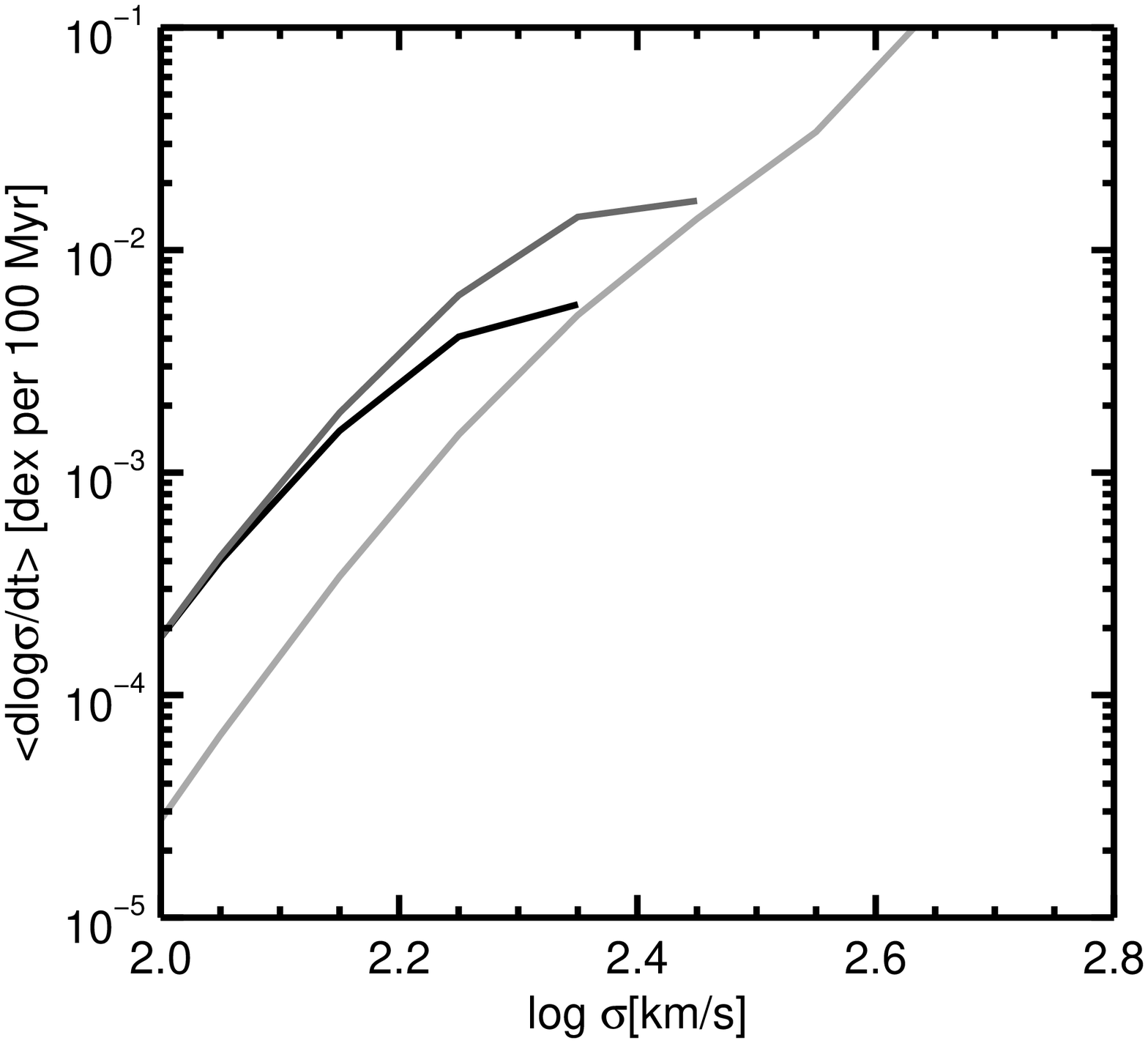} &
	\includegraphics[scale=0.35]{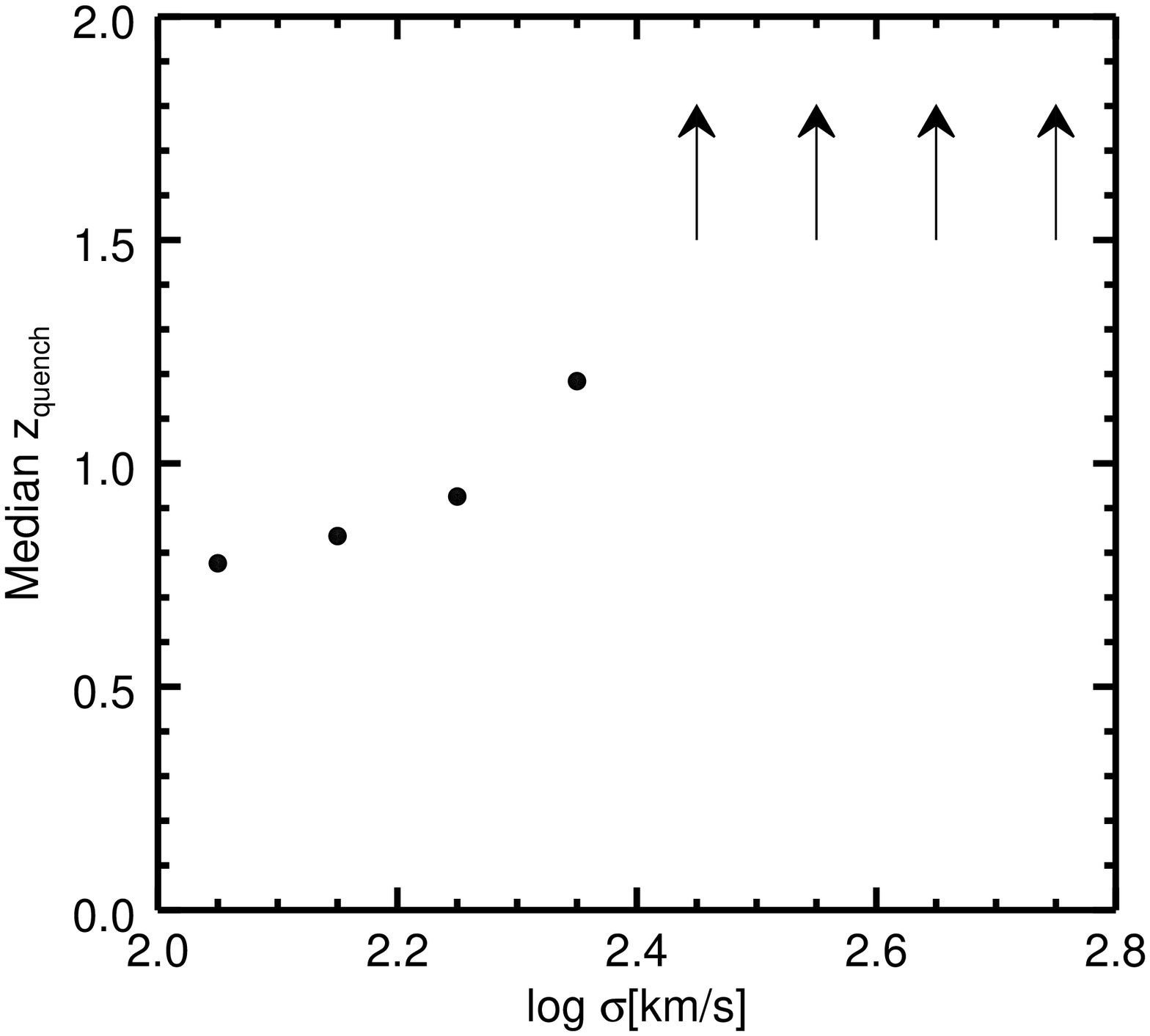} \\	
  \end{tabular}
    \caption{\emph{Top Left Panel:}  Fraction of newly quenched galaxies relative to the existing number of quiescent galaxies as a function of velocity dispersion.  At all redshifts, this fraction increases from zero evolution at the highest redshifts to $\sim0.1\unit{per 100 Myr}$ at low velocity dispersions.  \emph{Top Right Panel:}  Fraction of star-forming galaxies that are transformed into quiescent galaxies as a function of velocity dispersion.  This fraction exhibits the opposite behavior, with intermediate dispersion galaxies quenching much more rapidly than those with lower dispersions.  Again there is no evolution for galaxies with the highest dispersions.  \emph{Bottom Left Panel:} Rate of change in the velocity dispersions of star-forming galaxies as a function of velocity dispersion necessary to sustain the observed galaxy transformation and maintain a constant star-forming VDF.  This model requires extremely rapid changes in the velocity dispersions of star-forming galaxies, increasing with velocity dispersion.  \emph{Bottom Right Panel:} Median redshift at which galaxies quench in the simulation as a function of velocity dispersion.  In the model, high dispersion galaxies quench first and galaxies with lower velocity dispersions quench at later times.}
  \label{fig:simevolve_results}
 \end{figure*}

\begin{figure*}[!t]
  \centering
  	\plottwo{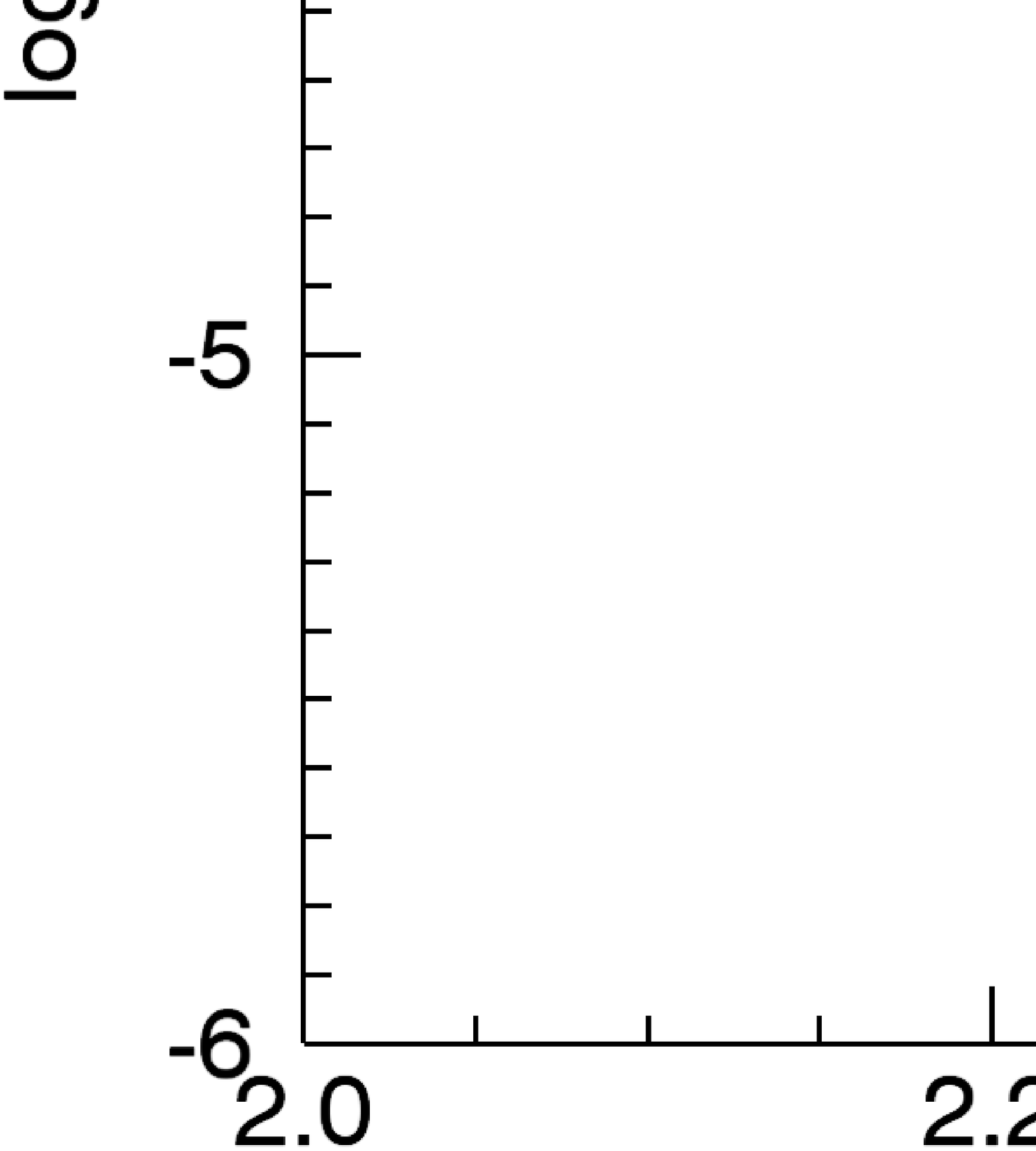}{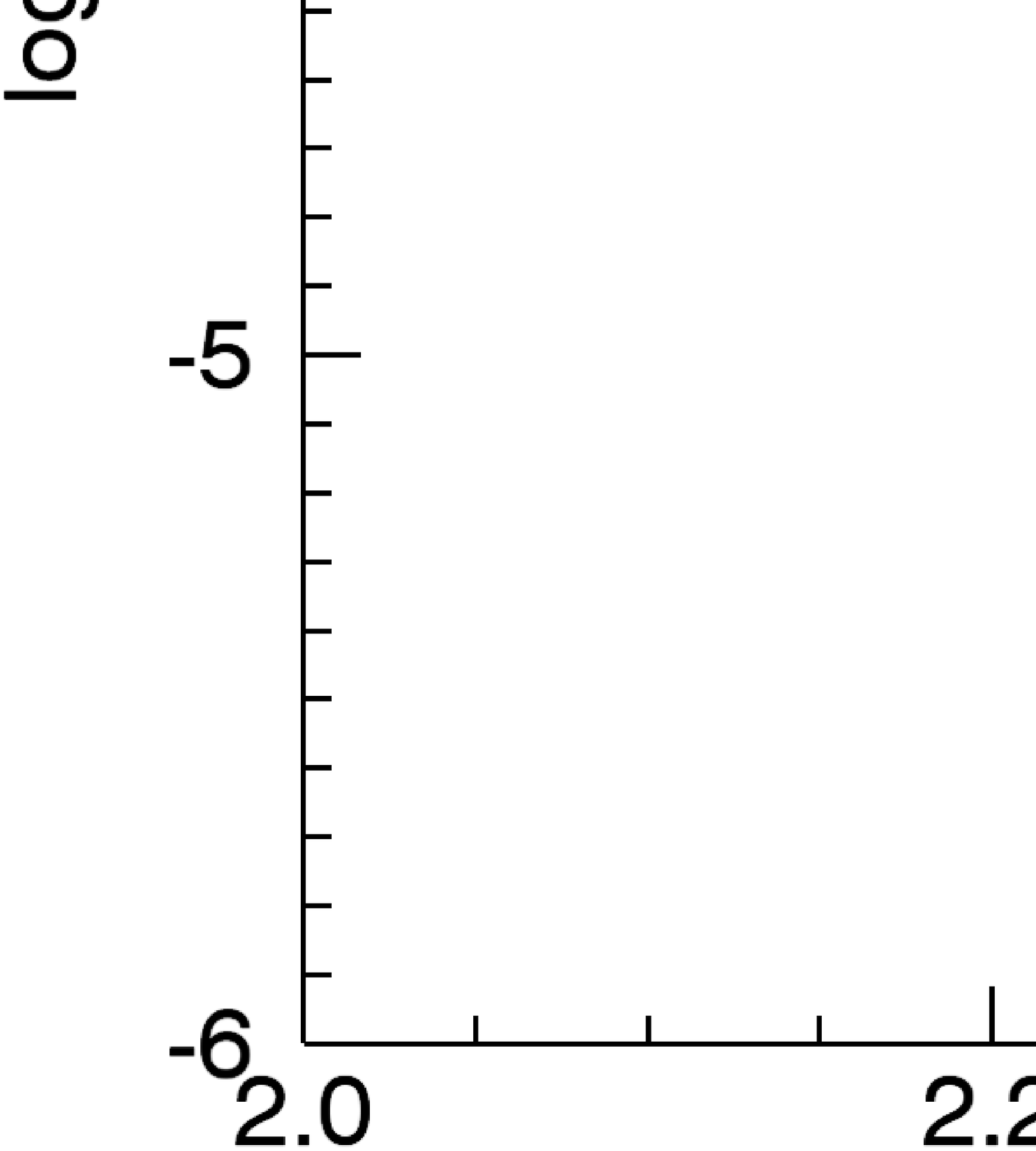}
    \caption{Evolution of the observed (large, transparent polygons) quiescent (\emph{Left Panel}) and star-forming (\emph{Right Panel}) VDFs and those from simple model (narrow, opaque, outlined polygons) including transformation and evolution in the velocity dispersions of individual star-forming galaxies.  Transformation rates are determined from the growth of the observed quiescent VDF.  In order to maintain the constant VDF for star-forming galaxies, $d\log\sigma/dt$ for each bin in velocity dispersion is tuned such that the number of galaxies increasing dispersions out of the bin matches the number of quenched galaxies.}
  \label{fig:simevolve}
 \end{figure*}

Figure \ref{fig:simevolve_results} demonstrates the overall properties required for the model to reproduce the observed VDF evolution.  In the left panel of Figure \ref{fig:simevolve_results} we present the implied growth fraction of quiescent galaxies as a function of velocity dispersion between each pair of redshift bins.  
\begin{equation}
f_{growth}(\sigma,z) = \frac{N_{q}(\sigma,z)-N_{q}(\sigma,z_i)}{N_{q}(\sigma,z)}
\end{equation}
As suggested by the evolution of the observed VDF, the growth in the number density of quiescent galaxies is strongest at lower inferred dispersions.  The opposite trend is exhibited by the quenching fraction (top right panel) of Figure \ref{fig:simevolve_results}.
\begin{equation}
f_{quench}(\sigma,z) = \frac{N_{q}(\sigma,z)-N_{q}(\sigma,z_i)}{N_{sf}(\sigma,z)}
\end{equation}
At the low velocity dispersions, star-forming galaxies are most common and the quenched fraction is low.  At high velocity dispersions, $\sigma\gtrsim 300\unit{km/s}$, the VDF does not evolve strongly and therefore the galaxies remain unchanged.  At intermediate velocity dispersions ($\sigma\sim250\unit{km/s}$), where few galaxies are quenching in a given timestep, the reservoir of star-forming galaxies is also minimal, therefore the quenched fraction is the highest at these velocity dispersions.  Finally, the rate at which star-forming galaxies enter into a bin of velocity dispersion implies an average rate of change in velocity dispersion, which is shown as a function of inferred dispersion in the bottom right panel of Figure  \ref{fig:simevolve_results}.  
\begin{equation}
\left<\frac{d\log\sigma}{dt}(\sigma,z)\right>= \Delta\sigma_{bin}\times\frac{f_{quench}(\sigma,z)}{dt}
\end{equation}
Here it is clear that the increasing quenching fraction with velocity dispersion implies that the rate at which the dispersions of galaxies must grow to maintain a constant distribution also increases with velocity dispersion.  This all combines to create a picture of transformations of intermediate dispersion star-forming galaxies, implying that at least in this regime, the quenching mechanism is rapid.   At the same time, some process (or set of processes) efficiently drives mass to or creates mass in the centers of galaxies more rapidly than growing their sizes, thus increasing their inferred velocity dispersions and replenishing the ranks of star-forming galaxies at a given velocity dispersion.

\subsection{Model Results and the ``Reservoir Problem''}

The results of this simulation are included in Figure \ref{fig:simevolve}, which demonstrates that the evolution of the quiescent and star-forming VDFs can be reproduced by this illustrative model.  The high velocity dispersion end of both VDFs is roughly constant in the simulation and there is a gradual build up of quiescent galaxies with low velocity dispersions.  As a result, quiescent galaxies with lower velocity dispersions will have the youngest stellar ages, which is consistent with observations of galaxies in the SDSS \citep[e.g.][]{haiman:07} and paints a complementary picture of the simulated build-up of the VDF for early type galaxies as presented in \citet{shankar:09}.  

It is clear that the high quiescent fraction and increasing number density of galaxies with intermediate velocity dispersions requires rapid quenching and structural evolution of star-forming galaxies.  While this result is quite extreme, it is hard to escape given the current observations.  One potential caveat is that there could be some selection bias such that current surveys of high redshift galaxies systematically missing obscured, star-forming and dynamically massive galaxies or galaxy centers which later evolve into compact quiescent galaxies.

We emphasize that this model by no means encompasses all physical processes that would in fact affect the evolution of the velocity dispersion functions.  Additional complications could include evolution in the dispersions of quiescent galaxies, perhaps due to the effects of minor merging, which has been shown to be important in decreasing the velocity dispersions of galaxies in cosmological simulations \citep{oser:11}.  Furthermore, major mergers could affect the velocity dispersion of quiescent galaxies, depending on the initial relative galaxy orbits and equal mass merging would decrease number densities.  However, while these effects might shuffle the distribution of quiescent galaxies, the qualitative agreement of our model should still apply.  Even with these possible complications, it appears that the VDFs require a strong build up of quiescent galaxies at low velocity dispersions combined with rapid quenching and increasing velocity dispersion for star-forming galaxies with intermediate velocity dispersions.

\section{Discussion and Conclusions}

In this paper, we investigate the evolution of the Velocity Dispersion Functions for star-forming and quiescent galaxies as estimated by photometric stellar masses and sizes.  We find three striking results:

\begin{itemize}
\item At the highest velocity dispersions ($\gtrsim 300 \unit{km/s}$), the number density of galaxies evolves very little with time if at all for both quiescent and star-forming galaxies.  This implies that galaxies with high velocity dispersions, like the halos they occupy, form at early times and, while they may evolve structurally, their potential wells and internal dynamics remain roughly unchanged with time.
\item The number density of quiescent galaxies with moderate to low velocity dispersions increases with time, presumably as star-forming galaxies with those dispersions shut off their star formation.
\item The VDF for star-forming galaxies remains roughly constant over the majority of cosmic time from $0.3<z\leq1.5$.  While there is potentially concern about the accuracy of inferred velocity dispersions of star-forming galaxies at $z>0$, these results also hold for the number density of star-forming galaxies as a function of surface density out to $z\sim1.2$.
\end{itemize}

We proposed a simple model to reproduce these effects including transformation of star-forming to quiescent galaxies and increasing velocity dispersions for star-forming galaxies.  In this case, galaxies with $\sigma \gtrsim 100 \unit{km/s}$ must quench rapidly, with an efficiency that increases with velocity dispersion up to $\sigma \sim 250 \unit{km/s}$ and rapid increases in velocity dispersions to maintain the distribution of star-forming galaxies.  

\citet{wake:12a} demonstrated that the velocity dispersion, and to a slightly lesser extent the surface density, of a galaxy is a good predictor of clustering and therefore possibly dark halo mass.  Therefore, if this result holds within the velocity dispersion range probed by this study and velocity dispersion remains a tied to halo mass out to $z\sim1.5$, we can speculate about the distribution of star-forming and quiescent galaxies as a function of velocity dispersion in the context of occupation of dark matter halos.  For example, given the stability of the distribution of velocity dispersions, or surface densities, for star-forming galaxies, these results would imply that the processes of accretion and gas cooling are tied to halo mass and at least in the most massive halos do not evolve with redshift, at least to $z\sim1.2$.  Given the evolution of the quiescent fraction, this would also suggest that star formation primarily turns off in lower mass halos at later times.  However this simple picture is complicated by the fact that the velocity dispersion (or size and stellar mass) of a galaxy can likely evolve more rapidly than the dark matter halo implies that this relation is likely to evolve with time, or at least exhibit a fair amount of scatter.  Furthermore, the fact that star-formation rates at a given stellar mass increase with redshift \cite[e.g.][]{daddi:07,noeske:07,whitaker:12b} contradicts a picture of constant gas accretion and cooling with time.

Overwhelming evidence has shown that every galaxy hosts a SMBH in its center and that the mass of that black hole is well correlated with the velocity dispersion of the host's bulge \citep[e.g.][]{magorrian:98,ferrarese:00,gebhardt:00,gultekin:09}.  Assuming that this correlation does not evolve strongly with time, the redshift evolution of the VDF implies that the build-up of SMBHs is anti-hierarchical; the most massive SMBHs are in place by $z\sim1.5$, whereas the smaller black holes are still growing at lower redshifts.  This ``down-sizing" is both consistent with other studies of the evolution of the SMBH function and with the idea that the most massive black holes are relics of the most active AGN activity, which peaked at $z\sim2$.

This work relies heavily on the accuracy of inferred velocity dispersions, which are calibrated locally.  Our conclusions are strengthened by the fact that the number density of galaxies as a function of surface density exhibits quite similar behavior.  However, all of this analysis relies on the accuracy of S\'{e}rsic profile fits and stellar mass estimates.  Size measurements are largely made from ground-based imaging and although \citet{bezanson:11} demonstrated the consistency of sizes in the NMBS COSMOS catalog with those measured from WFC3 imaging, the results would be less uncertain with the use of exclusively space-based imaging in the rest-frame optical, especially for the most compact galaxies.  Although corrections for non-homology via the $K_v(n)$ are small, repeating this study using only space-based imaging would vastly improve our ability accurately estimate best-fit S\'{e}rsic parameters.  Additionally, size measurements made with deeper imaging would allow us to probe to lower inferred dispersions at the higher redshifts probed by this study.  Stellar masses are based on a number of assumptions about the star formation histories of galaxies and IMF; we adopt a \cite{chabrier:03} IMF, however there have been hints that this could under-predict masses for the highest mass galaxies \citep[e.g.][]{dokkum:10wingford,cappellari:12}.  Finally, the accuracy of inferred velocity dispersion as a predictor of the actual dynamics of galaxies at $z>0$ remains an open question, especially for galaxies which are star-forming or with low velocity dispersions. This highlights the need for a more complete, statistical sample of measured velocity dispersions at $z>0$ to determine the accuracy of inferred velocity dispersions and evaluate the direct measurements of galaxy dynamics at $z>0$.

\begin{acknowledgements}
We would like to thank the referee for his/her thoughtful comments, Rik Williams, Ryan Quadri and Katherine Whitaker for providing access to the UDS and NMBS catalogs in electronic form as well as providing helpful feedback and David Wake for interesting discussions.  R.B. is grateful for the hospitality of Leiden University at which much of this work was performed. We also acknowledge financial support from HST grants GO-12177.01A and GO-12167.
\end{acknowledgements}

\appendix

\section{Incompleteness Limits in Velocity Dispersion}

\begin{figure}[h]
  \centering
    \centering
  	\begin{tabular}{cccc}
		\includegraphics[scale=0.45]{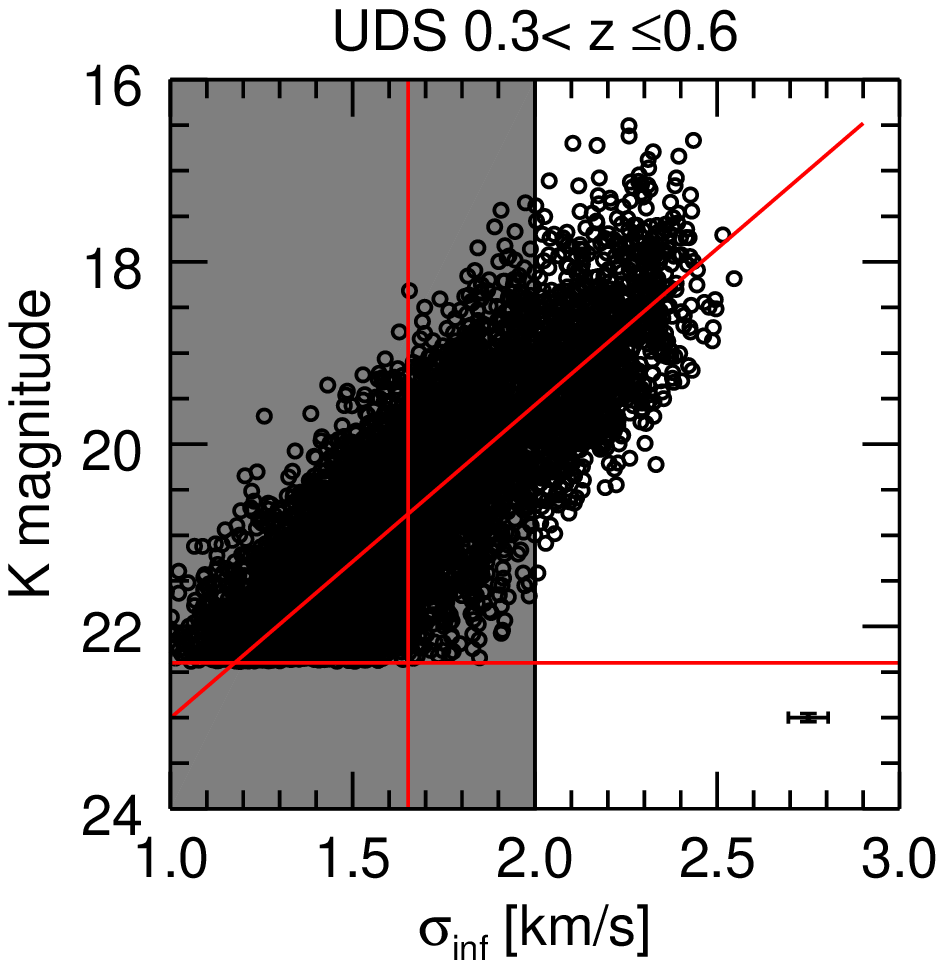} &
		\includegraphics[scale=0.45]{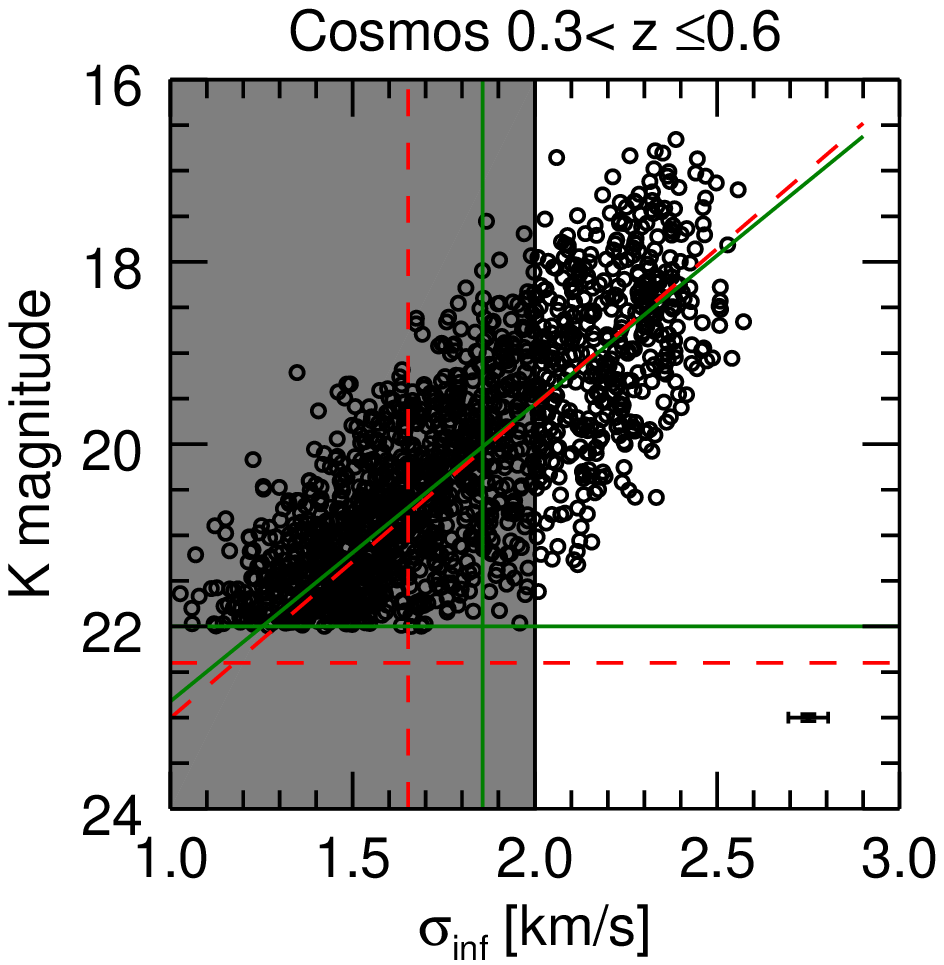} &
		\includegraphics[scale=0.45]{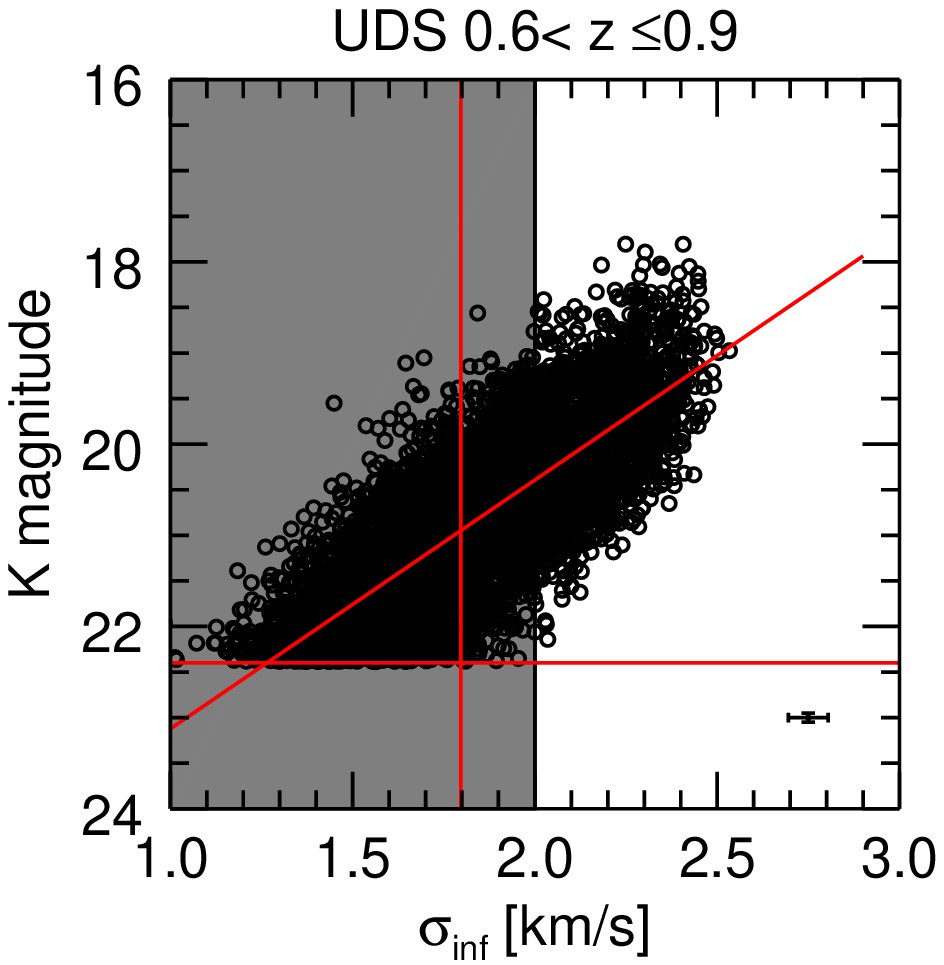} &
		\includegraphics[scale=0.45]{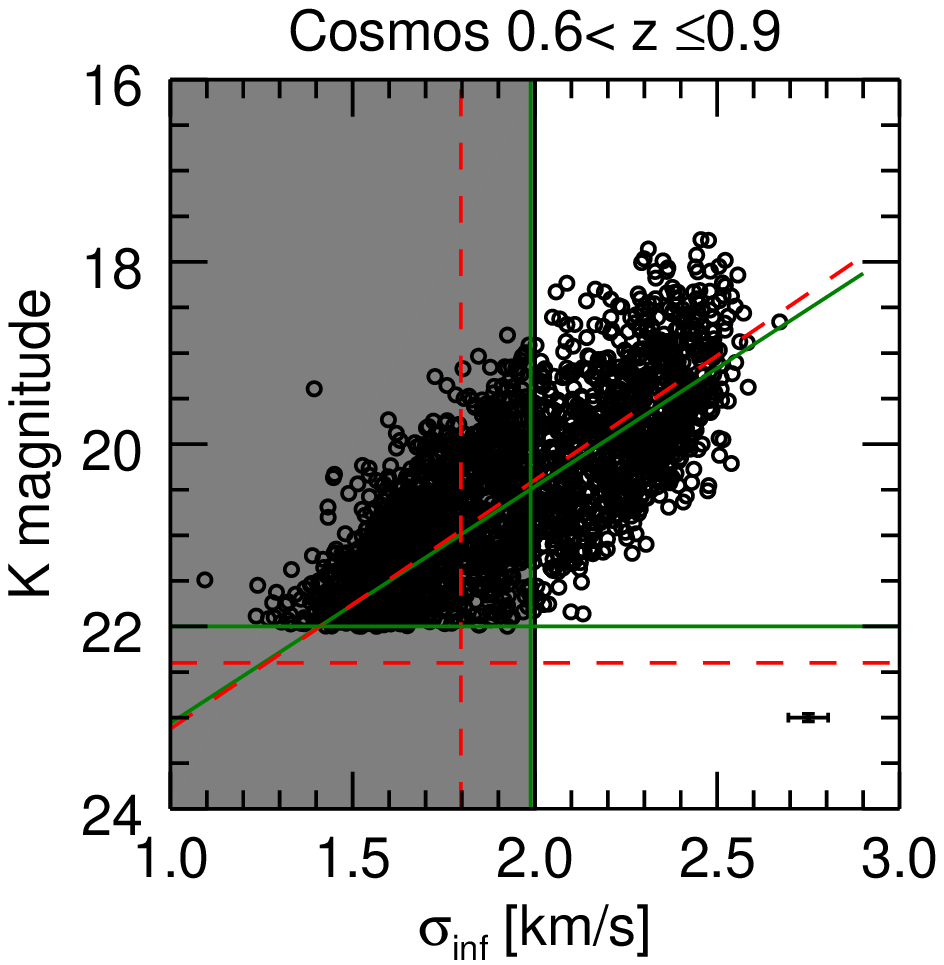} \\
		\includegraphics[scale=0.45]{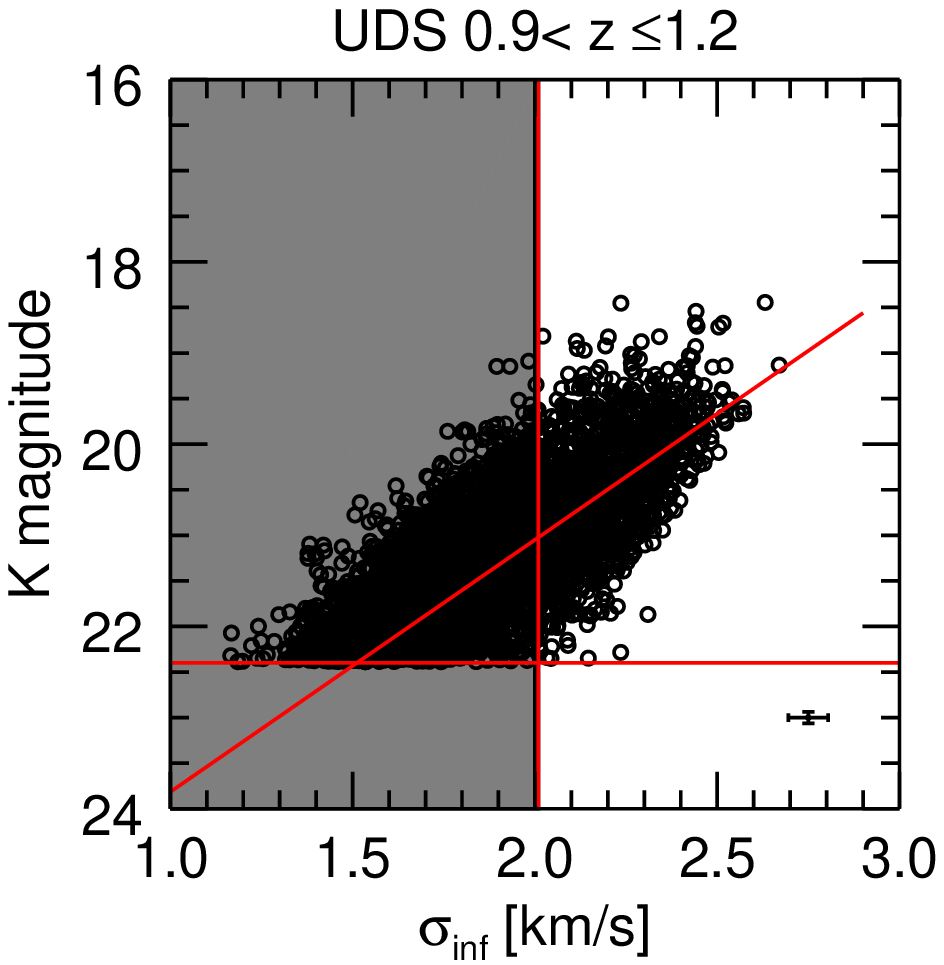} &
		\includegraphics[scale=0.45]{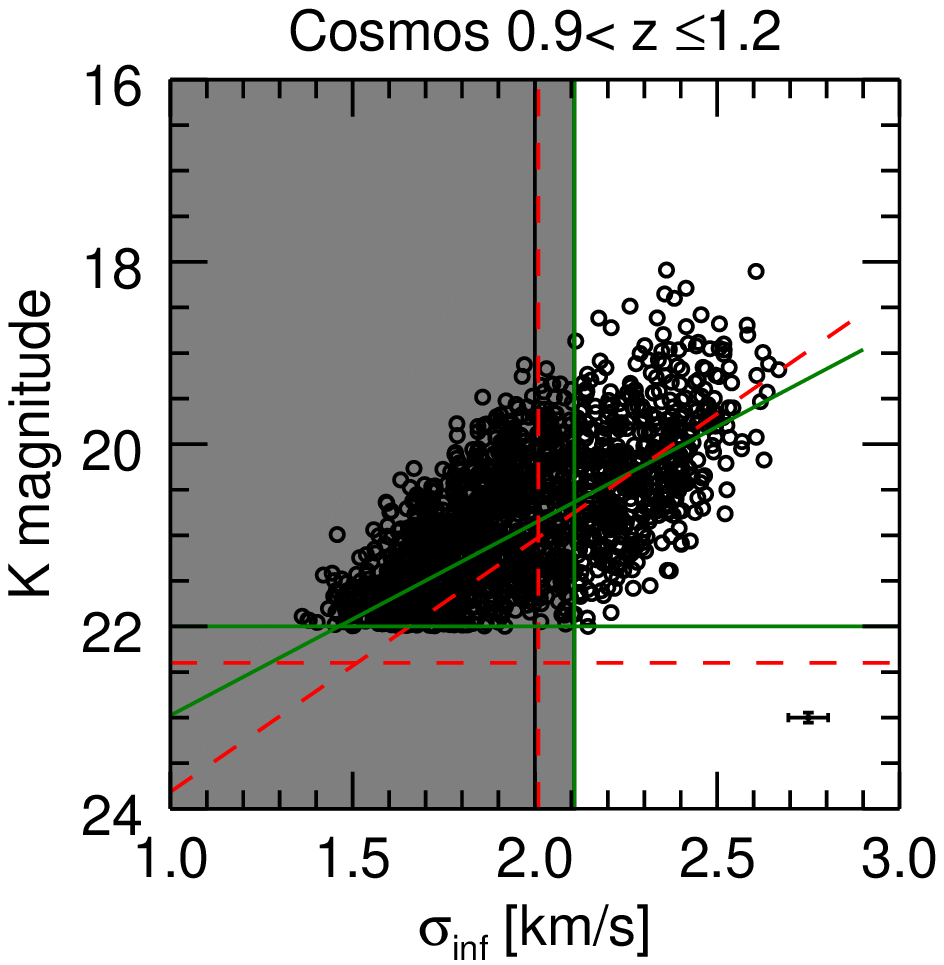} &
		\includegraphics[scale=0.45]{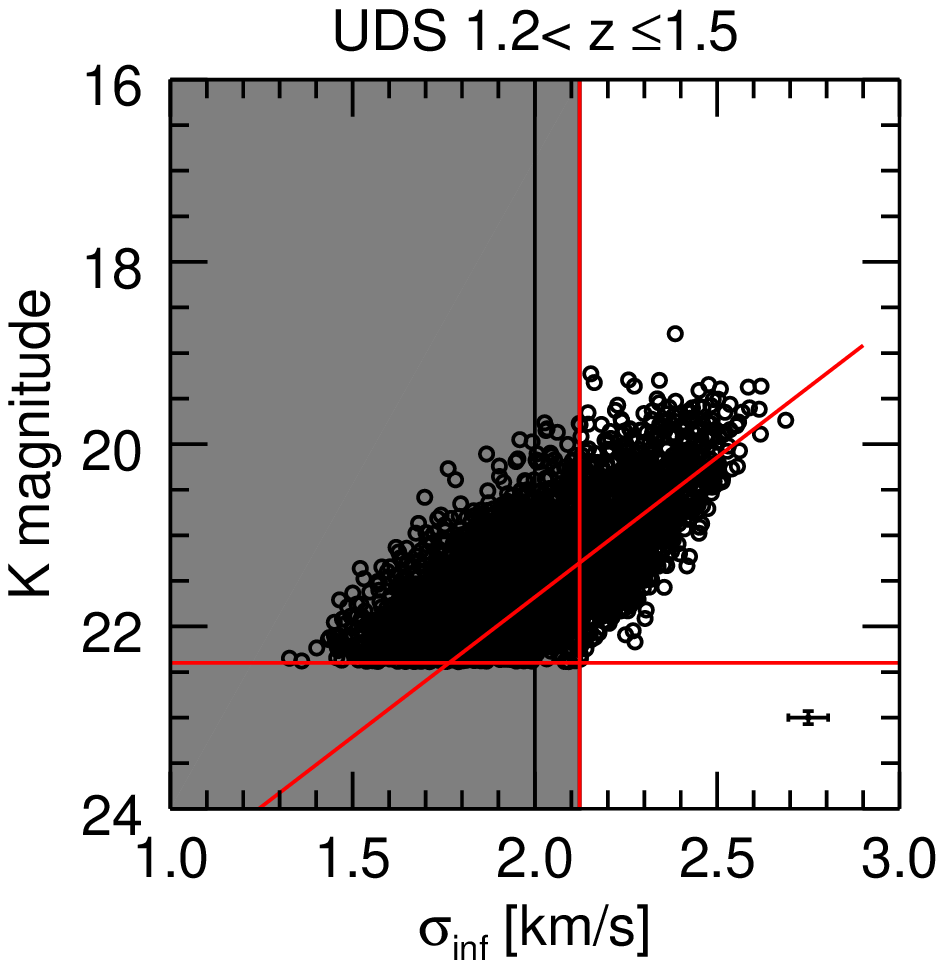} &
		\includegraphics[scale=0.45]{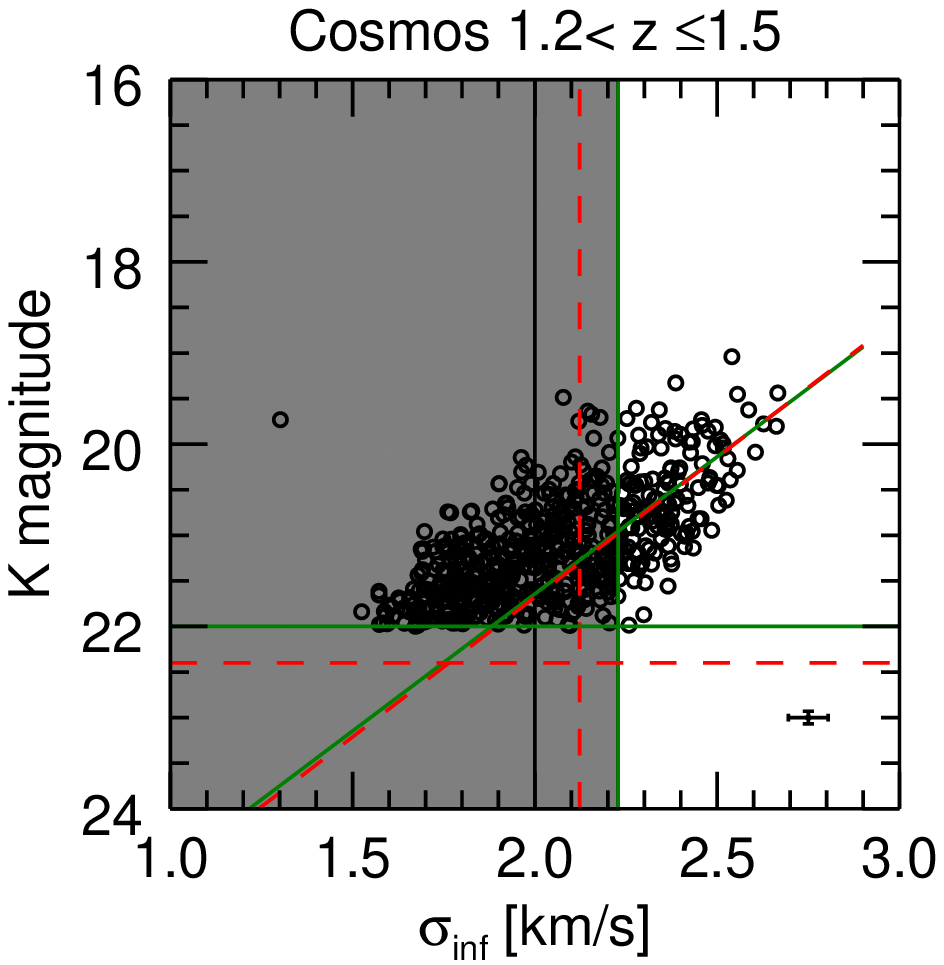} \\
  	\end{tabular}
    \caption{K magnitude vs. inferred velocity dispersion for UKIDSS UDS and NMBS Cosmos fields.  Black points represent galaxies in a given redshift range, horizontal lines reflect the magnitude limits of the size catalogs (Red for UDS at $K=22$, Green for Cosmos at $K=22.4$).  At each redshift incompleteness defined as the velocity dispersion at which twice the scatter about the overall linear trend (diagonal lines) meets the magnitude limit for the survey (horizontal lines).  Regions excluded from our measured VDFs, either due to incompleteness or the $\sigma_{inf}\ge100\unit{km/s}$ limit, are shaded in grey.  Median error bars are included in the lower right corner of each panel.}
      \label{fig:ksig}
 \end{figure}

Both the NMBS Cosmos and UKIDSS UDS photometric catalogs probe to remarkably faint magnitudes and therefore are complete to low stellar masses at all redshifts probed by this study ($\log M_{\star} \sim 9$ at the lowest redshifts and $\log M_{\star} \sim 10$ at $z\sim1.5$, however size measurements were only made for a bright subsample of galaxies in each survey.  For the NMBS Cosmos field, measurements down to a K magnitude limit of 22, for the UDS field $K \ge 22.4$.  Therefore, incompleteness in this study will be dominated by galaxies that are too faint to be included in the size catalogs.  In Figure \ref{fig:ksig} we show the K magnitude vs. inferred velocity dispersion for separately for each field and each redshift range.  In each case, galaxies fall around a generic trend (as indicated by diagonal lines: red-UDS, green-Cosmos) with a fair amount of scatter, which depends on velocity dispersion.  We fit the trend in each field for $\log\sigma > 2.0$.  While the trend evolves to brighter magnitudes at low redshift, the relations measured agree extremely well from field to field.  Assuming that these relations hold below the magnitude limits of our survey, we define conservative incompleteness limits as the velocity dispersion by calculating a $2-\sigma$ upper limit to the K magnitude for a given velocity dispersion and calculate the limiting $\sigma$, plus $0.06\unit{dex}$ for the scatter between $\sigma_0$ and $\sigma_{inf}$, at the magnitude limit for each redshift and field (indicated by vertical lines in Figure \ref{fig:ksig}).  Therefore, we shade regions in Figure \ref{fig:ksig} that are excluded from our analysis either due to the general $\log \sigma>2.0$ cut or our defined incompleteness limits.

\section{Redshift Distribution of Galaxies in the NMBS Cosmos and UKIDDS UDS Fields}

\begin{figure}[!t]
  \centering
	\includegraphics[scale=0.8]{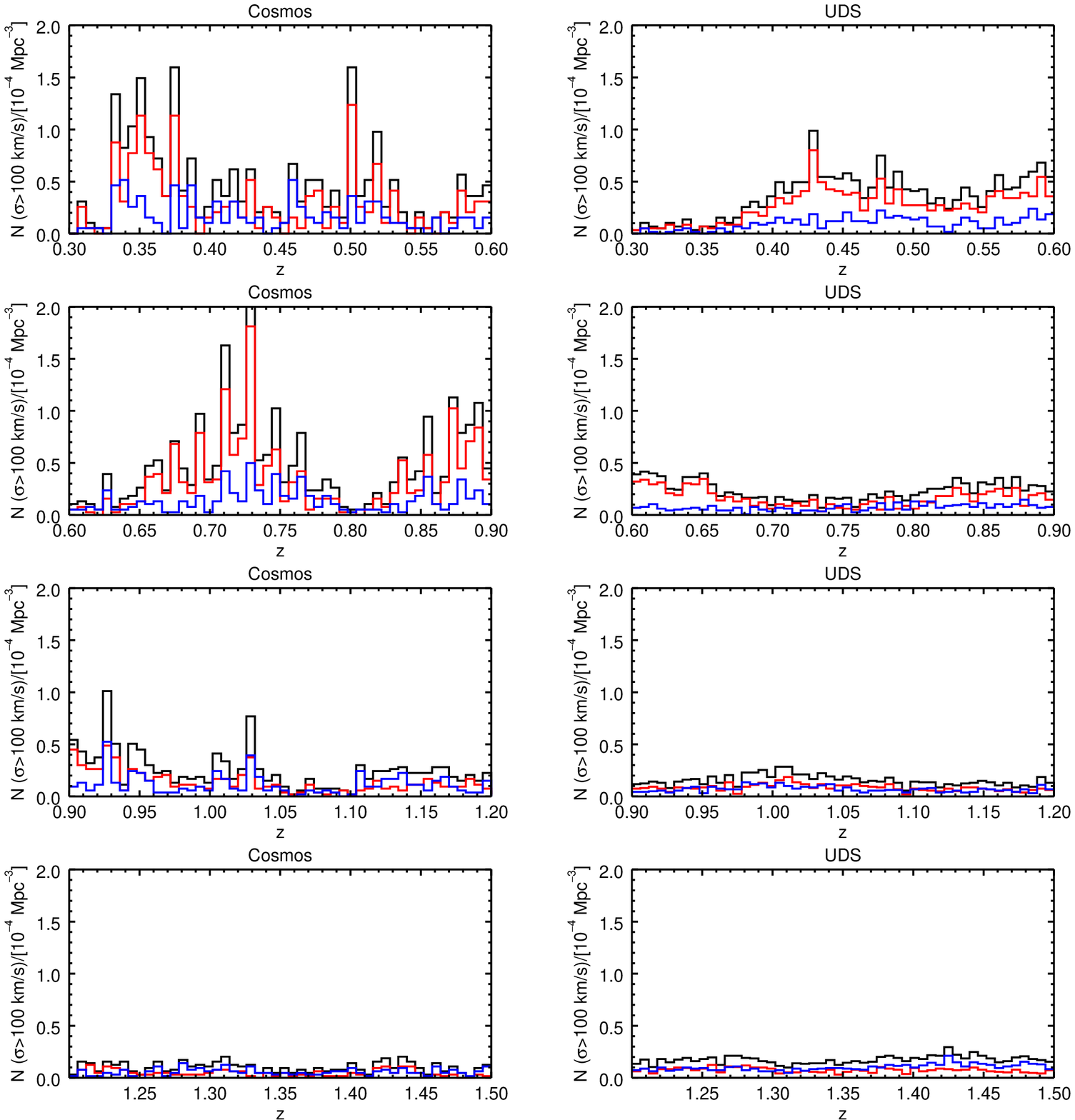}
    \caption{Redshift distributions of galaxies in the NMBS Cosmos and UKIDSS UDS fields.  Black histograms reflect the overall distribution of galaxies and blue/red histograms show the distribution of star-forming and quiescent galaxies respectively.}
  \label{fig:redshifts}
 \end{figure}

Simple estimates of the cosmic variance for these redshift bins predict much smaller discrepancies than are observed between the UDS and Cosmos surveys.  In Figure \ref{fig:redshifts} we present the redshift distributions of galaxies in these two surveys.   In particular we note the peaks in the Cosmos field, primarily the overdensity at $z\sim0.7$ and at $z\sim0.9$, which likely boost the number density of galaxies in the $0.6<z<0.9$ redshift bin.  

\section{Velocity Dispersion Functions and Error Estimates}

We include the measured velocity dispersion functions in tabular form with all calculated error estimates: Poisson errors ($\pm Poi$), cosmic variance ($\pm cv$), systematic redshift uncertainties ($\pm (z)$), systematic uncertainties in stellar mass measurements ($\pm (m)$) and the total error bars with all errors added in quadrature for the individual fields and volume-weighted combined measurements for quiescent galaxies ($\Phi_q$) and star-forming galaxies ($\Phi_{sf}$).
 
\begin{longtable*}[!t]{cccccccccccccc}
\tablecaption{Inferred Velocity Dispersion Functions\label{tbl:VDF}}
\tablehead{
\colhead{$\log\sigma_{inf}$} & \colhead{Field} & \colhead{$\log\Phi_q$} & \colhead{$\pm(Poi)$} & \colhead{$\pm(cv)$} & \colhead{$\pm(z)$} & \colhead{$\pm(m)$} & \colhead{$\pm(Tot)$} & \colhead{$\log\Phi_{sf}$} & \colhead{$\pm(Poi)$} & \colhead{$\pm(cv)$} &  \colhead{$\pm(z)$} & \colhead{$\pm(m)$} & \colhead{$\pm(Tot)$} \\ 
\colhead{$[km/s]$} & & \colhead{$[\unit{{Mpc}^{-3}{dex}^{-1}}]$} & & & & & & \colhead{$[\unit{{Mpc}^{-3} \rm{dex}^{-1}}]$} & & & & & }
\\ [-1ex]
\hline
$ 0.3 < z \leq 0.6$ \\*
\hline  \\ [+0.5ex]
$2.0-2.1$ \\* 
\nodata & \emph{Cosmos} & $-2.68$ & $0.07$ & $0.06$ & $^{+0.03}_{-0.03}$ & $^{+0.03}_{-0.00}$ & $^{+0.10}_{-0.09}$ & $-2.41$ & $0.05$ & $0.05$ & $^{+0.04}_{-0.09}$ & $^{+0.04}_{-0.10}$ & $^{+0.09}_{-0.15}$ \\* [+1.5ex]
\nodata & \emph{UDS} & $-2.53$ & $0.03$ & $0.04$ & $^{+0.13}_{-0.13}$ & $^{+0.13}_{-0.20}$ & $^{+0.19}_{-0.24}$ & $-2.56$ & $0.03$ & $0.05$ & $^{+0.12}_{-0.08}$ & $^{+0.09}_{-0.14}$ & $^{+0.16}_{-0.17}$ \\* [+1.5ex]
\nodata & \emph{Total} & $-2.56$ & $0.03$ & $0.04$ & $^{+0.12}_{-0.12}$ & $^{+0.12}_{-0.18}$ & $^{+0.18}_{-0.23}$ & $-2.52$ & $0.03$ & $0.04$ & $^{+0.10}_{-0.07}$ & $^{+0.08}_{-0.12}$ & $^{+0.14}_{-0.15}$ \\* 
$2.1-2.2$ \\* 
\nodata & \emph{Cosmos} & $-2.50$ & $0.06$ & $0.05$ & $^{+0.07}_{-0.06}$ & $^{+0.05}_{-0.14}$ & $^{+0.12}_{-0.17}$ & $-2.63$ & $0.06$ & $0.06$ & $^{+0.07}_{-0.04}$ & $^{+0.12}_{-0.16}$ & $^{+0.16}_{-0.18}$ \\* [+1.5ex]
\nodata & \emph{UDS} & $-2.39$ & $0.03$ & $0.04$ & $^{+0.09}_{-0.09}$ & $^{+0.00}_{-0.02}$ & $^{+0.10}_{-0.11}$ & $-2.87$ & $0.05$ & $0.05$ & $^{+0.14}_{-0.17}$ & $^{+0.16}_{-0.11}$ & $^{+0.23}_{-0.21}$ \\* [+1.5ex]
\nodata & \emph{Total} & $-2.42$ & $0.03$ & $0.04$ & $^{+0.08}_{-0.09}$ & $^{+0.00}_{-0.02}$ & $^{+0.09}_{-0.10}$ & $-2.80$ & $0.04$ & $0.04$ & $^{+0.12}_{-0.15}$ & $^{+0.14}_{-0.09}$ & $^{+0.19}_{-0.18}$ \\* 
$2.2-2.3$ \\* 
\nodata & \emph{Cosmos} & $-2.43$ & $0.05$ & $0.05$ & $^{+0.03}_{-0.00}$ & $^{+0.08}_{-0.02}$ & $^{+0.12}_{-0.08}$ & $-2.89$ & $0.09$ & $0.06$ & $^{+0.05}_{-0.00}$ & $^{+0.11}_{-0.25}$ & $^{+0.16}_{-0.27}$ \\* [+1.5ex]
\nodata & \emph{UDS} & $-2.45$ & $0.03$ & $0.04$ & $^{+0.13}_{-0.06}$ & $^{+0.04}_{-0.03}$ & $^{+0.15}_{-0.08}$ & $-3.24$ & $0.07$ & $0.06$ & $^{+0.01}_{-0.21}$ & $^{+0.26}_{-0.42}$ & $^{+0.28}_{-0.48}$ \\* [+1.5ex]
\nodata & \emph{Total} & $-2.44$ & $0.03$ & $0.04$ & $^{+0.12}_{-0.05}$ & $^{+0.03}_{-0.03}$ & $^{+0.13}_{-0.07}$ & $-3.12$ & $0.06$ & $0.05$ & $^{+0.01}_{-0.17}$ & $^{+0.21}_{-0.34}$ & $^{+0.23}_{-0.39}$ \\* 
$2.3-2.4$ \\* 
\nodata & \emph{Cosmos} & $-2.35$ & $0.05$ & $0.05$ & $^{+0.00}_{-0.06}$ & $^{+0.01}_{-0.23}$ & $^{+0.07}_{-0.24}$ & $-3.21$ & $0.13$ & $0.07$ & $^{+0.03}_{-0.08}$ & $^{+0.07}_{-0.18}$ & $^{+0.16}_{-0.24}$ \\* [+1.5ex]
\nodata & \emph{UDS} & $-2.65$ & $0.04$ & $0.05$ & $^{+0.06}_{-0.00}$ & $^{+0.17}_{-0.36}$ & $^{+0.19}_{-0.36}$ & $-3.77$ & $0.14$ & $0.07$ & $^{+0.00}_{-0.22}$ & $^{+0.11}_{-0.15}$ & $^{+0.19}_{-0.31}$ \\* [+1.5ex]
\nodata & \emph{Total} & $-2.55$ & $0.03$ & $0.04$ & $^{+0.05}_{-0.01}$ & $^{+0.14}_{-0.30}$ & $^{+0.16}_{-0.30}$ & $-3.55$ & $0.09$ & $0.05$ & $^{+0.01}_{-0.16}$ & $^{+0.08}_{-0.12}$ & $^{+0.14}_{-0.23}$ \\* 
$2.4-2.5$ \\* 
\nodata & \emph{Cosmos} & $-2.89$ & $0.09$ & $0.06$ & $^{+0.05}_{-0.00}$ & $^{+0.31}_{-0.17}$ & $^{+0.33}_{-0.20}$ & $-3.99$ & $0.31$ & $0.09$ & $^{+0.48}_{-0.30}$ & $^{+0.60}_{-0.30}$ & $^{+0.83}_{-0.53}$ \\* [+1.5ex]
\nodata & \emph{UDS} & $-3.34$ & $0.08$ & $0.06$ & $^{+0.07}_{-0.11}$ & $^{+0.33}_{-0.48}$ & $^{+0.36}_{-0.50}$ & $-4.77$ & $0.43$ & $0.11$ & $^{+0.30}_{-99.00}$ & $^{+0.85}_{-99.00}$ & $^{+1.00}_{-99.00}$ \\* [+1.5ex]
\nodata & \emph{Total} & $-3.18$ & $0.06$ & $0.05$ & $^{+0.06}_{-0.08}$ & $^{+0.26}_{-0.37}$ & $^{+0.28}_{-0.39}$ & $-4.42$ & $0.25$ & $0.07$ & $^{+0.26}_{-99.00}$ & $^{+0.56}_{-99.00}$ & $^{+0.69}_{-99.00}$ \\* 
$2.5-2.6$ \\* 
\nodata & \emph{Cosmos} & $-3.33$ & $0.14$ & $0.07$ & $^{+0.05}_{-0.65}$ & $^{+0.28}_{-0.65}$ & $^{+0.32}_{-0.94}$ & $<-4.29$ & \nodata & \nodata & \nodata & \nodata & \nodata \\* [+1.5ex]
\nodata & \emph{UDS} & $-4.47$ & $0.31$ & $0.09$ & $^{+0.00}_{-0.00}$ & $^{+0.65}_{-99.00}$ & $^{+0.73}_{-99.00}$ & $<-4.77$ & \nodata & \nodata & \nodata & \nodata & \nodata \\* [+1.5ex]
\nodata & \emph{Total} & $-3.85$ & $0.13$ & $0.06$ & $^{+0.03}_{-0.39}$ & $^{+0.31}_{-99.00}$ & $^{+0.35}_{-99.00}$ & $<-4.89$ & \nodata & \nodata & \nodata & \nodata & \nodata \\* 
$2.6-2.7$ \\* 
\nodata & \emph{Cosmos} & $<-4.29$ & \nodata & \nodata & \nodata & \nodata & \nodata & $<-4.29$ & \nodata & \nodata & \nodata & \nodata & \nodata \\ [+1ex]
\nodata & \emph{UDS} & $<-4.77$ & \nodata & \nodata & \nodata & \nodata & \nodata & $<-4.77$ & \nodata & \nodata & \nodata & \nodata & \nodata \\ [+1ex]
\nodata & \emph{Total} & $<-4.89$ & \nodata & \nodata & \nodata & \nodata & \nodata & $<-4.89$ & \nodata & \nodata & \nodata & \nodata & \nodata \\ [+1ex] 
\hline
$ 0.6 < z \leq 0.9$ \\*
\hline  \\ [+0.5ex]
$2.0-2.1$ \\* 
\nodata & \emph{Cosmos} & $-2.82$ & $0.06$ & $0.05$ & $^{+0.06}_{-0.00}$ & $^{+0.13}_{-0.06}$ & $^{+0.16}_{-0.10}$ & $-2.51$ & $0.04$ & $0.05$ & $^{+0.15}_{-0.19}$ & $^{+0.09}_{-0.10}$ & $^{+0.19}_{-0.23}$ \\* [+1.5ex]
\nodata & \emph{UDS} & $-2.79$ & $0.03$ & $0.04$ & $^{+0.00}_{-0.18}$ & $^{+0.06}_{-0.10}$ & $^{+0.08}_{-0.21}$ & $-2.71$ & $0.03$ & $0.04$ & $^{+0.05}_{-0.00}$ & $^{+0.11}_{-0.10}$ & $^{+0.13}_{-0.11}$ \\* [+1.5ex]
\nodata & \emph{Total} & $-2.79$ & $0.03$ & $0.04$ & $^{+0.01}_{-0.16}$ & $^{+0.05}_{-0.09}$ & $^{+0.07}_{-0.19}$ & $-2.65$ & $0.02$ & $0.03$ & $^{+0.05}_{-0.03}$ & $^{+0.10}_{-0.09}$ & $^{+0.12}_{-0.10}$ \\* 
$2.1-2.2$ \\* 
\nodata & \emph{Cosmos} & $-2.62$ & $0.05$ & $0.05$ & $^{+0.09}_{-0.03}$ & $^{+0.12}_{-0.07}$ & $^{+0.16}_{-0.10}$ & $-2.72$ & $0.05$ & $0.05$ & $^{+0.14}_{-0.09}$ & $^{+0.12}_{-0.08}$ & $^{+0.19}_{-0.14}$ \\* [+1.5ex]
\nodata & \emph{UDS} & $-2.71$ & $0.03$ & $0.04$ & $^{+0.06}_{-0.03}$ & $^{+0.02}_{-0.02}$ & $^{+0.08}_{-0.06}$ & $-2.93$ & $0.04$ & $0.04$ & $^{+0.05}_{-0.07}$ & $^{+0.13}_{-0.22}$ & $^{+0.15}_{-0.24}$ \\* [+1.5ex]
\nodata & \emph{Total} & $-2.69$ & $0.02$ & $0.04$ & $^{+0.05}_{-0.03}$ & $^{+0.02}_{-0.02}$ & $^{+0.07}_{-0.05}$ & $-2.87$ & $0.03$ & $0.04$ & $^{+0.05}_{-0.06}$ & $^{+0.11}_{-0.19}$ & $^{+0.13}_{-0.20}$ \\* 
$2.2-2.3$ \\* 
\nodata & \emph{Cosmos} & $-2.35$ & $0.03$ & $0.05$ & $^{+0.03}_{-0.09}$ & $^{+0.05}_{-0.15}$ & $^{+0.08}_{-0.18}$ & $-2.91$ & $0.06$ & $0.05$ & $^{+0.04}_{-0.08}$ & $^{+0.11}_{-0.09}$ & $^{+0.14}_{-0.15}$ \\* [+1.5ex]
\nodata & \emph{UDS} & $-2.60$ & $0.03$ & $0.04$ & $^{+0.00}_{-0.11}$ & $^{+0.00}_{-0.09}$ & $^{+0.05}_{-0.15}$ & $-3.39$ & $0.06$ & $0.05$ & $^{+0.09}_{-0.00}$ & $^{+0.24}_{-0.31}$ & $^{+0.27}_{-0.32}$ \\* [+1.5ex]
\nodata & \emph{Total} & $-2.52$ & $0.02$ & $0.03$ & $^{+0.00}_{-0.10}$ & $^{+0.01}_{-0.08}$ & $^{+0.04}_{-0.13}$ & $-3.21$ & $0.04$ & $0.04$ & $^{+0.07}_{-0.02}$ & $^{+0.18}_{-0.23}$ & $^{+0.20}_{-0.24}$ \\* 
$2.3-2.4$ \\* 
\nodata & \emph{Cosmos} & $-2.31$ & $0.03$ & $0.05$ & $^{+0.08}_{-0.11}$ & $^{+0.02}_{-0.05}$ & $^{+0.10}_{-0.13}$ & $-3.18$ & $0.09$ & $0.06$ & $^{+0.08}_{-0.06}$ & $^{+0.18}_{-0.40}$ & $^{+0.22}_{-0.42}$ \\* [+1.5ex]
\nodata & \emph{UDS} & $-2.79$ & $0.03$ & $0.04$ & $^{+0.00}_{-0.02}$ & $^{+0.15}_{-0.16}$ & $^{+0.16}_{-0.17}$ & $-4.06$ & $0.14$ & $0.07$ & $^{+0.15}_{-0.00}$ & $^{+0.36}_{-0.15}$ & $^{+0.42}_{-0.22}$ \\* [+1.5ex]
\nodata & \emph{Total} & $-2.61$ & $0.02$ & $0.03$ & $^{+0.02}_{-0.03}$ & $^{+0.11}_{-0.12}$ & $^{+0.12}_{-0.13}$ & $-3.64$ & $0.07$ & $0.05$ & $^{+0.09}_{-0.03}$ & $^{+0.21}_{-0.20}$ & $^{+0.25}_{-0.22}$ \\* 
$2.4-2.5$ \\* 
\nodata & \emph{Cosmos} & $-2.49$ & $0.04$ & $0.05$ & $^{+0.13}_{-0.22}$ & $^{+0.13}_{-0.22}$ & $^{+0.19}_{-0.32}$ & $-3.80$ & $0.18$ & $0.08$ & $^{+0.12}_{-0.30}$ & $^{+0.22}_{-0.30}$ & $^{+0.32}_{-0.47}$ \\* [+1.5ex]
\nodata & \emph{UDS} & $-3.32$ & $0.06$ & $0.05$ & $^{+0.05}_{-0.01}$ & $^{+0.37}_{-0.59}$ & $^{+0.38}_{-0.60}$ & $-4.58$ & $0.25$ & $0.08$ & $^{+0.00}_{-0.48}$ & $^{+0.37}_{-0.48}$ & $^{+0.45}_{-0.72}$ \\* [+1.5ex]
\nodata & \emph{Total} & $-2.94$ & $0.03$ & $0.04$ & $^{+0.06}_{-0.09}$ & $^{+0.22}_{-0.36}$ & $^{+0.24}_{-0.37}$ & $-4.23$ & $0.14$ & $0.06$ & $^{+0.05}_{-0.31}$ & $^{+0.24}_{-0.31}$ & $^{+0.30}_{-0.47}$ \\* 
$2.5-2.6$ \\* 
\nodata & \emph{Cosmos} & $-3.13$ & $0.08$ & $0.06$ & $^{+0.23}_{-0.45}$ & $^{+0.42}_{-0.67}$ & $^{+0.49}_{-0.81}$ & $<-4.58$ & \nodata & \nodata & \nodata & \nodata & \nodata \\* [+1.5ex]
\nodata & \emph{UDS} & $-4.76$ & $0.31$ & $0.09$ & $^{+0.30}_{-0.00}$ & $^{+0.85}_{-99.00}$ & $^{+0.95}_{-99.00}$ & $<-5.06$ & \nodata & \nodata & \nodata & \nodata & \nodata \\* [+1.5ex]
\nodata & \emph{Total} & $-3.71$ & $0.08$ & $0.05$ & $^{+0.20}_{-0.37}$ & $^{+0.37}_{-99.00}$ & $^{+0.44}_{-99.00}$ & $<-5.19$ & \nodata & \nodata & \nodata & \nodata & \nodata \\* 
$2.6-2.7$ \\* 
\nodata & \emph{Cosmos} & $-4.58$ & $0.43$ & $0.11$ & $^{+0.60}_{-99.00}$ & $^{+0.78}_{-0.00}$ & $^{+1.08}_{-99.00}$ & $<-4.58$ & \nodata & \nodata & \nodata & \nodata & \nodata \\ [+1ex]
\nodata & \emph{UDS} & $<-5.06$ & \nodata & \nodata & \nodata & \nodata & \nodata & $<-5.06$ & \nodata & \nodata & \nodata & \nodata & \nodata \\ [+1ex]
\nodata & \emph{Total} & $-5.19$ & $0.43$ & $0.11$ & $^{+0.60}_{-99.00}$ & $^{+0.78}_{-0.00}$ & $^{+1.08}_{-99.00}$ & $<-5.19$ & \nodata & \nodata & \nodata & \nodata & \nodata \\ [+1ex] 
\hline
$ 0.9 < z \leq 1.2$ \\*
\hline  \\ [+0.5ex]
$2.0-2.1$ \\* 
\nodata & \emph{Cosmos} & $-3.21$ & $0.08$ & $0.06$ & $^{+0.19}_{-0.31}$ & $^{+0.14}_{-0.10}$ & $^{+0.26}_{-0.34}$ & $-2.60$ & $0.04$ & $0.05$ & $^{+0.08}_{-0.12}$ & $^{+0.07}_{-0.06}$ & $^{+0.12}_{-0.15}$ \\* [+1.5ex]
\nodata & \emph{UDS} & $-3.21$ & $0.04$ & $0.04$ & $^{+0.07}_{-0.01}$ & $^{+0.06}_{-0.09}$ & $^{+0.11}_{-0.11}$ & $-2.74$ & $0.03$ & $0.04$ & $^{+0.04}_{-0.00}$ & $^{+0.11}_{-0.13}$ & $^{+0.13}_{-0.14}$ \\* [+1.5ex]
\nodata & \emph{Total} & $-3.21$ & $0.04$ & $0.04$ & $^{+0.07}_{-0.03}$ & $^{+0.06}_{-0.08}$ & $^{+0.11}_{-0.11}$ & $-2.70$ & $0.02$ & $0.03$ & $^{+0.04}_{-0.02}$ & $^{+0.10}_{-0.11}$ & $^{+0.11}_{-0.12}$ \\* 
$2.1-2.2$ \\* 
\nodata & \emph{Cosmos} & $-2.92$ & $0.05$ & $0.05$ & $^{+0.17}_{-0.18}$ & $^{+0.11}_{-0.14}$ & $^{+0.22}_{-0.24}$ & $-2.80$ & $0.05$ & $0.05$ & $^{+0.01}_{-0.00}$ & $^{+0.15}_{-0.13}$ & $^{+0.16}_{-0.15}$ \\* [+1.5ex]
\nodata & \emph{UDS} & $-3.03$ & $0.04$ & $0.04$ & $^{+0.08}_{-0.00}$ & $^{+0.09}_{-0.12}$ & $^{+0.13}_{-0.13}$ & $-2.96$ & $0.03$ & $0.04$ & $^{+0.07}_{-0.00}$ & $^{+0.09}_{-0.13}$ & $^{+0.13}_{-0.14}$ \\* [+1.5ex]
\nodata & \emph{Total} & $-3.00$ & $0.03$ & $0.04$ & $^{+0.07}_{-0.02}$ & $^{+0.08}_{-0.10}$ & $^{+0.12}_{-0.12}$ & $-2.92$ & $0.03$ & $0.03$ & $^{+0.06}_{-0.00}$ & $^{+0.08}_{-0.11}$ & $^{+0.11}_{-0.12}$ \\* 
$2.2-2.3$ \\* 
\nodata & \emph{Cosmos} & $-2.81$ & $0.05$ & $0.05$ & $^{+0.16}_{-0.17}$ & $^{+0.00}_{-0.01}$ & $^{+0.17}_{-0.18}$ & $-3.01$ & $0.06$ & $0.05$ & $^{+0.05}_{-0.12}$ & $^{+0.08}_{-0.15}$ & $^{+0.12}_{-0.21}$ \\* [+1.5ex]
\nodata & \emph{UDS} & $-2.94$ & $0.03$ & $0.04$ & $^{+0.04}_{-0.00}$ & $^{+0.00}_{-0.00}$ & $^{+0.07}_{-0.05}$ & $-3.30$ & $0.05$ & $0.05$ & $^{+0.05}_{-0.00}$ & $^{+0.21}_{-0.31}$ & $^{+0.22}_{-0.31}$ \\* [+1.5ex]
\nodata & \emph{Total} & $-2.91$ & $0.03$ & $0.03$ & $^{+0.04}_{-0.02}$ & $^{+0.00}_{-0.00}$ & $^{+0.06}_{-0.05}$ & $-3.21$ & $0.04$ & $0.04$ & $^{+0.04}_{-0.02}$ & $^{+0.17}_{-0.25}$ & $^{+0.19}_{-0.26}$ \\* 
$2.3-2.4$ \\* 
\nodata & \emph{Cosmos} & $-2.73$ & $0.04$ & $0.05$ & $^{+0.11}_{-0.12}$ & $^{+0.00}_{-0.09}$ & $^{+0.13}_{-0.17}$ & $-3.37$ & $0.09$ & $0.06$ & $^{+0.09}_{-0.00}$ & $^{+0.21}_{-0.13}$ & $^{+0.25}_{-0.17}$ \\* [+1.5ex]
\nodata & \emph{UDS} & $-2.98$ & $0.03$ & $0.04$ & $^{+0.06}_{-0.01}$ & $^{+0.04}_{-0.16}$ & $^{+0.09}_{-0.17}$ & $-3.98$ & $0.11$ & $0.06$ & $^{+0.30}_{-0.00}$ & $^{+0.37}_{-0.33}$ & $^{+0.49}_{-0.35}$ \\* [+1.5ex]
\nodata & \emph{Total} & $-2.90$ & $0.03$ & $0.03$ & $^{+0.05}_{-0.02}$ & $^{+0.03}_{-0.14}$ & $^{+0.08}_{-0.15}$ & $-3.73$ & $0.07$ & $0.05$ & $^{+0.21}_{-0.00}$ & $^{+0.26}_{-0.23}$ & $^{+0.35}_{-0.25}$ \\* 
$2.4-2.5$ \\* 
\nodata & \emph{Cosmos} & $-2.90$ & $0.05$ & $0.05$ & $^{+0.04}_{-0.09}$ & $^{+0.14}_{-0.15}$ & $^{+0.16}_{-0.19}$ & $-3.82$ & $0.15$ & $0.08$ & $^{+0.00}_{-0.43}$ & $^{+0.33}_{-0.90}$ & $^{+0.37}_{-1.01}$ \\* [+1.5ex]
\nodata & \emph{UDS} & $-3.41$ & $0.05$ & $0.05$ & $^{+0.10}_{-0.00}$ & $^{+0.27}_{-0.35}$ & $^{+0.29}_{-0.36}$ & $-4.91$ & $0.31$ & $0.09$ & $^{+0.54}_{-0.00}$ & $^{+0.60}_{-0.30}$ & $^{+0.87}_{-0.44}$ \\* [+1.5ex]
\nodata & \emph{Total} & $-3.22$ & $0.04$ & $0.04$ & $^{+0.07}_{-0.02}$ & $^{+0.20}_{-0.26}$ & $^{+0.22}_{-0.27}$ & $-4.33$ & $0.14$ & $0.06$ & $^{+0.23}_{-0.24}$ & $^{+0.32}_{-0.53}$ & $^{+0.43}_{-0.61}$ \\* 
$2.5-2.6$ \\* 
\nodata & \emph{Cosmos} & $-3.43$ & $0.10$ & $0.06$ & $^{+0.10}_{-0.00}$ & $^{+0.37}_{-0.22}$ & $^{+0.40}_{-0.25}$ & $<-4.73$ & \nodata & \nodata & \nodata & \nodata & \nodata \\* [+1.5ex]
\nodata & \emph{UDS} & $-4.09$ & $0.12$ & $0.06$ & $^{+0.00}_{-0.16}$ & $^{+0.33}_{-0.51}$ & $^{+0.36}_{-0.55}$ & $-5.21$ & $0.43$ & $0.10$ & $^{+0.00}_{-99.00}$ & $^{+0.00}_{-99.00}$ & $^{+0.45}_{-99.00}$ \\* [+1.5ex]
\nodata & \emph{Total} & $-3.81$ & $0.08$ & $0.05$ & $^{+0.03}_{-0.11}$ & $^{+0.25}_{-0.35}$ & $^{+0.27}_{-0.38}$ & $-5.33$ & $0.43$ & $0.10$ & $^{+0.00}_{-99.00}$ & $^{+0.00}_{-99.00}$ & $^{+0.45}_{-99.00}$ \\* 
$2.6-2.7$ \\* 
\nodata & \emph{Cosmos} & $-3.77$ & $0.14$ & $0.07$ & $^{+0.00}_{-0.65}$ & $^{+0.12}_{-0.95}$ & $^{+0.20}_{-1.17}$ & $<-4.73$ & \nodata & \nodata & \nodata & \nodata & \nodata \\ [+1ex]
\nodata & \emph{UDS} & $-4.91$ & $0.31$ & $0.09$ & $^{+0.00}_{-0.00}$ & $^{+0.30}_{-0.30}$ & $^{+0.44}_{-0.44}$ & $<-5.21$ & \nodata & \nodata & \nodata & \nodata & \nodata \\ [+1ex]
\nodata & \emph{Total} & $-4.29$ & $0.13$ & $0.06$ & $^{+0.00}_{-0.39}$ & $^{+0.14}_{-0.58}$ & $^{+0.21}_{-0.72}$ & $<-5.33$ & \nodata & \nodata & \nodata & \nodata & \nodata \\ [+1ex] 
\hline
$ 1.2 < z \leq 1.5$ \\*
\hline  \\ [+0.5ex]
$2.0-2.1$ \\* 
\nodata & \emph{Cosmos} & $-4.11$ & $0.19$ & $0.08$ & $^{+0.00}_{-0.10}$ & $^{+0.34}_{-0.10}$ & $^{+0.40}_{-0.25}$ & $-2.95$ & $0.05$ & $0.05$ & $^{+0.02}_{-0.01}$ & $^{+0.02}_{-0.03}$ & $^{+0.08}_{-0.08}$ \\* [+1.5ex]
\nodata & \emph{UDS} & $-3.51$ & $0.06$ & $0.05$ & $^{+0.04}_{-0.01}$ & $^{+0.22}_{-0.19}$ & $^{+0.23}_{-0.20}$ & $-2.71$ & $0.02$ & $0.03$ & $^{+0.00}_{-0.07}$ & $^{+0.03}_{-0.05}$ & $^{+0.05}_{-0.09}$ \\* [+1.5ex]
\nodata & \emph{Total} & $-3.60$ & $0.05$ & $0.05$ & $^{+0.04}_{-0.01}$ & $^{+0.21}_{-0.18}$ & $^{+0.23}_{-0.20}$ & $-2.76$ & $0.02$ & $0.03$ & $^{+0.00}_{-0.06}$ & $^{+0.03}_{-0.05}$ & $^{+0.05}_{-0.09}$ \\* 
$2.1-2.2$ \\* 
\nodata & \emph{Cosmos} & $-3.41$ & $0.09$ & $0.06$ & $^{+0.00}_{-0.02}$ & $^{+0.06}_{-0.36}$ & $^{+0.12}_{-0.37}$ & $-3.04$ & $0.06$ & $0.05$ & $^{+0.08}_{-0.00}$ & $^{+0.12}_{-0.16}$ & $^{+0.16}_{-0.18}$ \\* [+1.5ex]
\nodata & \emph{UDS} & $-3.12$ & $0.04$ & $0.04$ & $^{+0.00}_{-0.10}$ & $^{+0.08}_{-0.17}$ & $^{+0.10}_{-0.21}$ & $-2.81$ & $0.03$ & $0.04$ & $^{+0.00}_{-0.07}$ & $^{+0.05}_{-0.05}$ & $^{+0.07}_{-0.09}$ \\* [+1.5ex]
\nodata & \emph{Total} & $-3.18$ & $0.03$ & $0.04$ & $^{+0.00}_{-0.09}$ & $^{+0.08}_{-0.16}$ & $^{+0.09}_{-0.20}$ & $-2.86$ & $0.02$ & $0.03$ & $^{+0.00}_{-0.06}$ & $^{+0.05}_{-0.05}$ & $^{+0.07}_{-0.09}$ \\* 
$2.2-2.3$ \\* 
\nodata & \emph{Cosmos} & $-3.27$ & $0.07$ & $0.06$ & $^{+0.01}_{-0.13}$ & $^{+0.05}_{-0.07}$ & $^{+0.11}_{-0.18}$ & $-3.29$ & $0.08$ & $0.06$ & $^{+0.00}_{-0.07}$ & $^{+0.08}_{-0.16}$ & $^{+0.13}_{-0.20}$ \\* [+1.5ex]
\nodata & \emph{UDS} & $-3.02$ & $0.03$ & $0.04$ & $^{+0.00}_{-0.05}$ & $^{+0.01}_{-0.02}$ & $^{+0.05}_{-0.07}$ & $-3.00$ & $0.03$ & $0.04$ & $^{+0.06}_{-0.09}$ & $^{+0.14}_{-0.24}$ & $^{+0.16}_{-0.26}$ \\* [+1.5ex]
\nodata & \emph{Total} & $-3.07$ & $0.03$ & $0.04$ & $^{+0.00}_{-0.04}$ & $^{+0.01}_{-0.02}$ & $^{+0.05}_{-0.07}$ & $-3.06$ & $0.03$ & $0.04$ & $^{+0.06}_{-0.08}$ & $^{+0.13}_{-0.23}$ & $^{+0.15}_{-0.24}$ \\* 
$2.3-2.4$ \\* 
\nodata & \emph{Cosmos} & $-3.20$ & $0.07$ & $0.06$ & $^{+0.08}_{-0.00}$ & $^{+0.00}_{-0.03}$ & $^{+0.12}_{-0.10}$ & $-3.63$ & $0.11$ & $0.07$ & $^{+0.12}_{-0.00}$ & $^{+0.19}_{-0.33}$ & $^{+0.26}_{-0.36}$ \\* [+1.5ex]
\nodata & \emph{UDS} & $-3.03$ & $0.03$ & $0.04$ & $^{+0.00}_{-0.06}$ & $^{+0.03}_{-0.07}$ & $^{+0.06}_{-0.11}$ & $-3.38$ & $0.05$ & $0.04$ & $^{+0.03}_{-0.05}$ & $^{+0.14}_{-0.18}$ & $^{+0.15}_{-0.19}$ \\* [+1.5ex]
\nodata & \emph{Total} & $-3.07$ & $0.03$ & $0.04$ & $^{+0.01}_{-0.06}$ & $^{+0.02}_{-0.07}$ & $^{+0.05}_{-0.10}$ & $-3.43$ & $0.04$ & $0.04$ & $^{+0.03}_{-0.05}$ & $^{+0.13}_{-0.17}$ & $^{+0.15}_{-0.18}$ \\* 
$2.4-2.5$ \\* 
\nodata & \emph{Cosmos} & $-3.41$ & $0.09$ & $0.06$ & $^{+0.23}_{-0.06}$ & $^{+0.17}_{-0.12}$ & $^{+0.30}_{-0.17}$ & $-4.51$ & $0.31$ & $0.10$ & $^{+0.48}_{-0.00}$ & $^{+0.54}_{-0.00}$ & $^{+0.79}_{-0.32}$ \\* [+1.5ex]
\nodata & \emph{UDS} & $-3.24$ & $0.04$ & $0.04$ & $^{+0.00}_{-0.09}$ & $^{+0.13}_{-0.26}$ & $^{+0.15}_{-0.29}$ & $-3.99$ & $0.10$ & $0.06$ & $^{+0.11}_{-0.05}$ & $^{+0.43}_{-0.60}$ & $^{+0.46}_{-0.61}$ \\* [+1.5ex]
\nodata & \emph{Total} & $-3.27$ & $0.04$ & $0.04$ & $^{+0.02}_{-0.09}$ & $^{+0.12}_{-0.25}$ & $^{+0.14}_{-0.27}$ & $-4.07$ & $0.09$ & $0.06$ & $^{+0.11}_{-0.04}$ & $^{+0.42}_{-0.58}$ & $^{+0.45}_{-0.59}$ \\* 
$2.5-2.6$ \\* 
\nodata & \emph{Cosmos} & $-3.76$ & $0.13$ & $0.07$ & $^{+0.16}_{-0.00}$ & $^{+0.24}_{-0.34}$ & $^{+0.32}_{-0.37}$ & $-4.81$ & $0.43$ & $0.11$ & $^{+0.00}_{-0.00}$ & $^{+0.30}_{-99.00}$ & $^{+0.54}_{-99.00}$ \\* [+1.5ex]
\nodata & \emph{UDS} & $-3.94$ & $0.09$ & $0.06$ & $^{+0.00}_{-0.02}$ & $^{+0.44}_{-0.44}$ & $^{+0.46}_{-0.45}$ & $-4.99$ & $0.31$ & $0.09$ & $^{+0.00}_{-99.00}$ & $^{+0.40}_{-0.00}$ & $^{+0.51}_{-99.00}$ \\* [+1.5ex]
\nodata & \emph{Total} & $-3.89$ & $0.08$ & $0.05$ & $^{+0.02}_{-0.02}$ & $^{+0.38}_{-0.38}$ & $^{+0.39}_{-0.39}$ & $-4.93$ & $0.25$ & $0.08$ & $^{+0.00}_{-99.00}$ & $^{+0.34}_{-99.00}$ & $^{+0.44}_{-99.00}$ \\* 
$2.6-2.7$ \\* 
\nodata & \emph{Cosmos} & $-4.20$ & $0.22$ & $0.09$ & $^{+0.10}_{-0.60}$ & $^{+0.10}_{-0.30}$ & $^{+0.27}_{-0.71}$ & $<-4.81$ & \nodata & \nodata & \nodata & \nodata & \nodata \\ [+1ex]
\nodata & \emph{UDS} & $-4.99$ & $0.31$ & $0.09$ & $^{+0.00}_{-0.30}$ & $^{+0.60}_{-0.30}$ & $^{+0.68}_{-0.53}$ & $-4.99$ & $0.31$ & $0.09$ & $^{+0.00}_{-0.30}$ & $^{+0.00}_{-99.00}$ & $^{+0.32}_{-99.00}$ \\ [+1ex]
\nodata & \emph{Total} & $-4.63$ & $0.18$ & $0.06$ & $^{+0.04}_{-0.30}$ & $^{+0.36}_{-0.22}$ & $^{+0.42}_{-0.43}$ & $-5.11$ & $0.31$ & $0.09$ & $^{+0.00}_{-0.30}$ & $^{+0.00}_{-99.00}$ & $^{+0.32}_{-99.00}$ \\ [+1ex] 
\end{longtable*}

\section{Velocity Dispersion Function for Star-Forming Galaxies - Including Gas Masses}

\begin{figure*}[]
  \centering
	\includegraphics[scale=0.2]{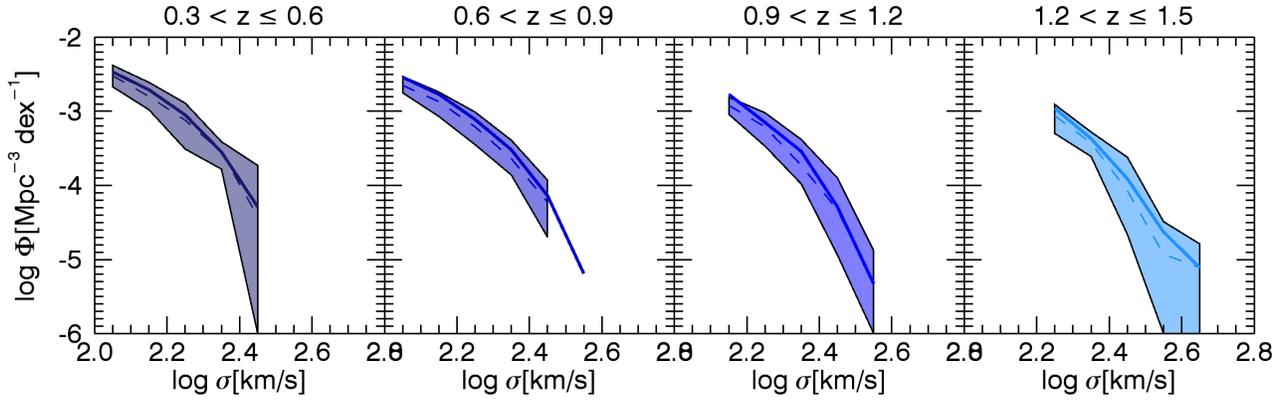}
    \caption{Velocity dispersion function for star-forming galaxies calculated without (dashed lines and shaded polygons) and with (solid lines) inclusion of gas mass.}
  \label{fig:VDF_gas}
 \end{figure*}

As discussed in \S2.2, increased specific star formation rates at $z\gtrsim1$ imply that galaxies at high redshifts must have elevated gas surface densities and therefore it may be important to include gas masses in addition to $M_{\star}$ in calculating inferred velocity dispersions.  For each galaxy, we calculate the inferred velocity dispersion using $M_{baryon}$, as defined in \S2.2 and recalculate the VDF.  This has very little effect on the VDF for quiescent galaxies, but it does increase the velocity dispersions of star-forming galaxies, slightly shifting the number density distribution (by at most $\sim10\%$).  In Figure \ref{fig:VDF_gas} we show the VDF for star-forming galaxies with (solid line) and without the inclusion of gas.  These shifts are within the quoted error bars for the VDF and given uncertainties in calculated sSFRs, we ignore this effect for the analysis in this Paper.  We expect the effect of gas to be increasingly important at higher redshifts than probed by this study.  

%\bibliography{/Users/rachel/Documents/Papers/all}

\end{document}